\documentclass[
reprint,
superscriptaddress,
twocolumn,
showpacs,
amsmath,amssymb,amsthm,amsfonts, 
aps,
prl,
floatfix,notitlepage,latexsym
]{revtex4-2}

\usepackage{graphicx}
\usepackage{comment,xcolor}

\usepackage{bm}
\usepackage{braket}
\usepackage{comment}
\usepackage{ulem}

\newcommand{\lsec}[1]{\textit{#1.---}}

\usepackage{hyperref} 
\definecolor{lightblue}{RGB}{73,151,208}
\definecolor{crimson}{RGB}{140,41,53}
\hypersetup{
    colorlinks,
    linkcolor={crimson},
    citecolor={lightblue},
    urlcolor={lightblue}
}

\begin{document}
\title{Robust effective ground state in a nonintegrable Floquet quantum circuit}
\author{Tatsuhiko N. Ikeda}
\affiliation{RIKEN Center for Quantum Computing, Wako, Saitama 351-0198, Japan}
\affiliation{Department of Physics, Boston University, Boston, Massachusetts 02215, USA}
\author{Sho Sugiura}
\affiliation{Physics and Informatics Laboratory, NTT Research, Inc., Sunnyvale, California, 94085, USA}
\affiliation{Laboratory for Nuclear Science, Massachusetts Institute of Technology, Cambridge, 02139, MA, USA}
\author{Anatoli Polkovnikov}
\affiliation{Department of Physics, Boston University, Boston, Massachusetts 02215, USA}
\date{\today}

\begin{abstract}
An external periodic (Floquet) drive is believed to bring any initial state to the featureless infinite temperature state in generic nonintegrable isolated quantum many-body systems in the thermodynamic limit, irrespective of the driving frequency $\Omega$. However, numerical or analytical evidence either proving or disproving this hypothesis is very limited and the issue has remained unsettled.
Here, we study the initial state dependence of Floquet heating in a nonintegrable kicked Ising chain of length up to $L=30$ with an efficient quantum circuit simulator, showing a possible counterexample: The ground state of the effective Floquet Hamiltonian is exceptionally robust against heating, and could stay at finite energy density even after infinitely many Floquet cycles, if the driving period is shorter than a threshold value. This sharp energy localization transition/crossover does not happen for generic excited states. The exceptional robustness of the ground state is interpreted by (i) its isolation in the energy spectrum and (ii) the fact that those states with $L$-independent $\hbar\Omega$ energy above the ground state energy of any generic local Hamiltonian, like the approximate Floquet Hamiltonian, are atypical and viewed as a collection of noninteracting quasipartiles. Our finding paves the way for 
engineering Floquet protocols with finite driving periods realizing long-lived, or possibly even perpetual, Floquet phases by initial state design.
\end{abstract}
\maketitle

\lsec{Introduction}
Periodically driven, or Floquet, quantum systems have recently attracted renewed attention from the viewpoint of Floquet engineering, i.e., creating intriguing functionalities of matter with external periodic drives~\cite{Goldman2014,Bukov2015,Holthaus2015,Eckardt2017,Oka2019}, together with rapid developments of experimental techniques, such as strong light-matter interactions and driven artificial quantum matter~\cite{Wang2013,Rechtsman2013,Jotzu2014,Choi2017,Zhang2017}.
In isolated systems Floquet-engineered states are believed to break down eventually due to heating~\cite{Lazarides2014,DAlessio2014,Kim2014}, i.e., the energy injection accompanied by the drive, and stability of Floquet engineering has been a central issue.
For general local Hamiltonians with bounded local energy spectrum, rigorous upper bounds on heating are known and guarantee that the heating is suppressed exponentially in the driving frequency $\Omega$ irrespective of the initial states~\cite{Kuwahara2016,Abanin2017,Avdoshkin2020}.
Many experimental~\cite{Rubio-Abadal2020,Viebahn2021,Peng2021,Beatrez2021} and numerical~\cite{Bukov2016heating,Machado2019,Machado2020,Luitz2020,Pizzi2020,Ye2020,Yin2021,Fleckenstein2021a,Fleckenstein2021b,Ikeda2021} studies observe actual heating rates obeying the exponential scaling consistent with these bounds in generic Hamiltonians with some notable exceptions~\cite{Prosen_1998,DAlessio2013b, Else2016,Yao2017,Heyl2019, Medenjak2020, Haldar2021,Emonts2022}). At the same time it is known that these bounds cannot be tight. For example, exponential heating was also observed in classical systems, where these bounds diverge due to the infinite local Hilbert-space size~\cite{Rajak_2018,Howell2019}.

At the same time, some numerical studies report indications of very sharp phase transition-like behavior of heating when the driving frequency is varied~\cite{Prosen2007,DAlessio2013b,Haldar2018,Sieberer2019,Kargi2021,Vernier2023}.
Namely, below (above) a threshold frequency, the system remains at a finite (is brought to the infinite) temperature after many driving cycles.
This sharp transition has also been translated to Trotterization on digital quantum computers, and the long-time Trotter error, a counterpart of heating, has been discussed~\cite{Heyl2019,Sieberer2019,Kargi2021}.
Yet those results cannot be conclusively extrapolated to the thermodynamic limit because of potentially large finite-size effects.
In a one-body chaotic model~\cite{Sieberer2019} and a special integrable model~\cite{Vernier2023}, such a Trotter transition has been analytically obtained.
On the other hand, some studies report smooth crossovers rather than a transition in generic nonintegrable models~\cite{DAlessio2014}.
Those studies differ in many ways, including the model, initial states, physical observables, etc., and it has yet to be understood whether and in what sense a phase transition exists in generic nonintegrable many-body models.

\begin{figure}[ht]
         \includegraphics[width=0.9\columnwidth]{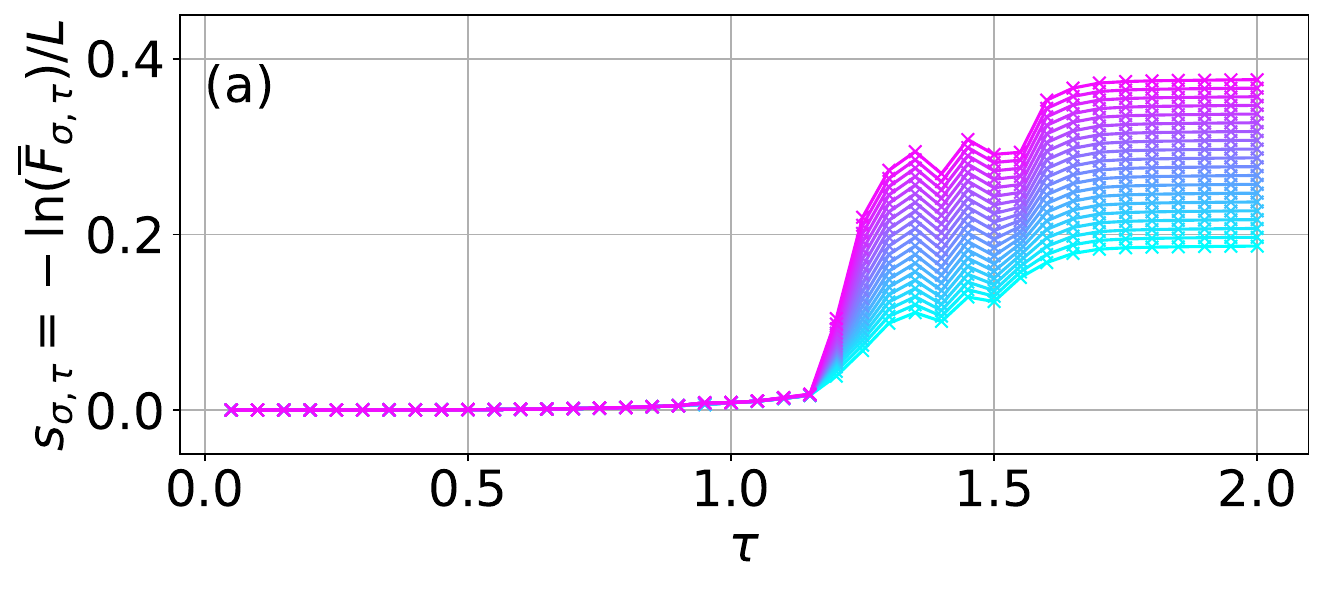}
         \includegraphics[width=0.9\columnwidth]{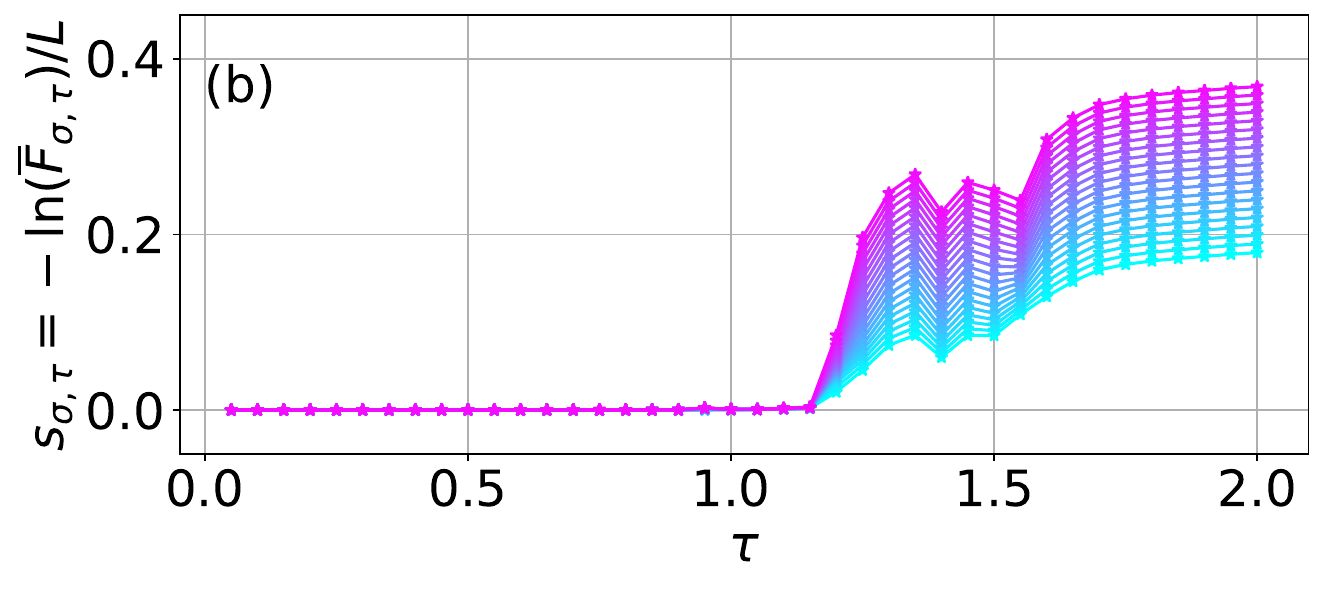}
	\caption{
        Time-averaged fidelity~\eqref{eq:time_averaged_fidelity_sigma} in the log scale under the Floquet evolution~\eqref{eq:Ht} starting from the ground states of (a) $H_F^{(0)}$ and (b) $H_F^{(2)}$.
        The system size is $L=24$, and plot colors from light blue to magenta correspond to the time cutoff $\sigma$ from $10^2$ to $10^4$ equidistant in the log scale.
        }
	\label{fig:transition}
\end{figure}

In this Letter, we show numerical evidence for the Trotter (or heating) transition in a nonintegrable kicked Ising model by reaching as large as $L=30$ spins with an efficient quantum circuit simulator (see Fig.~\ref{fig:transition}).  
The sharp transition is absent for most initial states but is most conspicuously seen when the initial state is the effective ground state, i.e., the ground state of an effective Hamiltonian in the high-frequency expansion.  It becomes sharper and sharper if we increase the order of the expansion, but the transition point is insensitive to this order.
We interpret the exceptional robustness of the effective ground state by the generic property of local Hamiltonians that the energetically resonant states with the ground state above energy $\hbar\Omega=O(1)$ consist of a few quasiparticles behaving freely in the thermodynamic limit.
The initial state dependence of the transition sheds new light on the seemingly contradicting previous reports about the presence/absence of transitions.
Besides, the stability of states above a critical drive frequency (i.e., below a critical Trotter step) encourages Floquet engineering (Trotter simulations) for a long time even in the thermodynamic limit without the need of scaling the Trotter step down to zero with increasing the simulation time~\cite{Heyl2019}.

\lsec{Formulation of the problem}
We consider a quantum spin-1/2 chain of length $L$ under the following time-periodic Hamiltonian
\begin{align}\label{eq:Ht}
    H(t) = \begin{cases}
        H_1 & \left( k\le t/\tau < k+\frac{1}{2} \right) \\
        H_2 & \left( k+\frac{1}{2} \le t/\tau < k+\frac{3}{2} \right)\\
        H_1 & \left(  k+\frac{3}{2} \le t/\tau < k+2 \right),
    \end{cases}
\end{align}
where $k\in\mathbb{Z}$ and $2\tau$ is the driving period, and
\begin{align}
    H_1 &= -\sum_{j=1}^L \left(\frac{J}{4} \sigma^z_j \sigma^z_{j+1}+\frac{h}{2}\sigma^z_j \right),\
    H_2 = -\frac{g}{2}\sum_{j=1}^L  \sigma^x_j.
\end{align}
Here $\sigma^x_j$ and $\sigma^z_j$ are the Pauli matrices acting on the site $j$, and the periodic boundary conditions are imposed while Ref.~\cite{Heyl2019} used the open ones.
Throughout this paper, we set $J=h=g=1$ since we have confirmed that the results are not sensitive to their choice as long as they are far away from integrable points.
An initial state $\ket{\psi_\mathrm{ini}}$ unitarily evolves in time under $H(t)$, and the state at $t=n\tau$ $(n\in \mathbb{Z})$ is given by $\ket{\psi(n\tau)}=T(\tau)^n \ket{\psi_\mathrm{ini}}$ with
\begin{align}\label{eq:Trotter}
    T(\tau)&=e^{-iH_1\frac{\tau}{2}}e^{-i H_2\tau}e^{-iH_1\frac{\tau}{2}}.
\end{align}

We ask now how stable $\ket{\psi(n\tau)}$ is under the periodic drive.
To quantify the stability, we introduce the following fidelity~\cite{ODea2023}
\begin{align}\label{eq:def_fidelity}
    F(n,\tau) = |\braket{\psi_\mathrm{ini}| T(\tau)^n|\psi_\mathrm{ini}}|^2.
\end{align}
This definition is motivated by the following reasoning.
The Magnus expansion (or the symmetric Baker--Campbell--Hausdorff formula) gives us a power series expansion for $H_F$ defined through $T(\tau)=e^{-i H_F \tau}$ as $H_F=\sum_{l=0}^\infty h^{(2l)}\tau^{2l}$, where odd-order terms all vanish due to the symmetry $T(-\tau)=T^\dag(\tau)$.
Then a truncated effective Hamiltonian $H_F^{(2k)}=\sum_{l=0}^k h^{(2l)}\tau^{2l}$ gives an approximation $U_{2k}(\tau)=e^{-iH_F^{(2k)}\tau}=T(\tau)+O(\tau^{2k+3})$.
The approximate unitary $U_{2k}(\tau)$ is generated by the time-independent Hamiltonian $H_F^{(2k)}$ and thus energy conserving, and the time evolution $\ket{\psi_{2k}(n\tau)}=U_{2k}(n\tau)\ket{\psi_\mathrm{ini}}$ is free from Floquet heating.
If $\ket{\psi_\mathrm{ini}}$ is an eigenstate of $H_F^{(2k)}$ (as we will assume below), the fidelity between the exact and approximate states $|\braket{\psi_{2k}(n\tau)|\psi(n\tau)}|^2$~\cite{Sieberer2019,Kargi2021} reduces to Eq.~\eqref{eq:def_fidelity}.
While Ref.~\cite{ODea2023} showed that eigenstates are more stable than superposition of them, we address which of the eigenstates are more stable.
We note that the above argument is also translated to Trotterization; $T(\tau)$ is a $(2k+2)$-th order Trotter approximation for $U_{2k}(n\tau)$ generated by the target Hamiltonian $H_F^{(2k)}$.
To focus on the long-time stability, we introduce the long- but finite-time average of the fidelity:
\begin{align}\label{eq:time_averaged_fidelity_sigma}
\overline{F}_{\sigma,\tau}=\frac{1}{\mathcal{N}_\sigma}\sum_{n=0}^\infty F(n,\tau)e^{-(n/\sigma)^2},
\end{align}
where $\mathcal{N}_\sigma:=\sum_{n=0}^\infty e^{-(n/\sigma)^2}$ and $\sigma$ $(>0)$ denotes a Gaussian cutoff.
The time-averaged fidelity is numerically obtained by calculating $T(\tau)^n\ket{\psi_\mathrm{ini}}$ for $n=0,1,\dots,n_\mathrm{max}$ so that $n_\mathrm{max}\gg \sigma$.
Since $T(\tau)$ can be represented by 1- and 2-qubit quantum gates unlike other Floquet models~\cite{DAlessio2014}, $T(\tau)^n\ket{\psi_\mathrm{ini}}$ is more efficiently calculated using a circuit simulator~\cite{Suzuki2021}.

\lsec{Sharp Trotter transition for the Floquet ground state}
Figure~\ref{fig:transition} shows the $\tau$-dependence of the time-averaged fidelity when the initial state is what we call here the effective Floquet ground state, i.e., the ground state of $H_F^{(0)}=H_1+H_2$ and $H_F^{(2)}$.
We remark that the Floquet ground state is not an eigenstate of $T(\tau)$ and hence $F(n,\tau)$ evolves in $n$.
For convenience, we plot the rate function
\begin{align}\label{eq:def_s}
s_{\sigma,\tau} = - L^{-1}\ln(\overline{F}_{\sigma,\tau}), 
\end{align}
which is expected to have a well-defined thermodynamic limit assuming the typical system size scaling $\overline{F}_{\sigma,\tau}=e^{-s_{\sigma,\tau} L}$ and a larger (smaller) $s_{\sigma,\tau}$ corresponding to a smaller (larger) fidelity.
If the unitary $T(\tau)$ can be represented as $T(\tau)=e^{-i H_F \tau}$ with $H_F$ being a local gapped Hamiltonian then, $s_{\sigma,\tau}$ is expected to be a small number well defined in the joint limit $\sigma\to \infty$, $L\to\infty$.
As Fig.~\ref{fig:transition}(a) shows, for $\tau<1.2$, $s_{\sigma,\tau}$ smoothly depends on $\tau$ and is almost independent of $\sigma$, whereas it increases abruptly in $\tau\ge1.2$ and simultaneously acquires strong $\sigma$-dependence. Thus the GS of $H_F^{(0)}$ is, at least, strongly robust against heating for $\tau< 1.2$.
This transition-like behavior at $\tau\approx1.2$ becomes more conspicuous when the initial state is the ground state of $H_F^{(2)}$, for which
the fidelity remains almost constant all the way to the transition point $\tau=1.2$.
Note that this behavior is similar to what was obtained in an integrable circuit~\cite{Vernier2023} for a special initial state, whereas our circuit is nonintegrable.

\begin{figure}
         \includegraphics[width=0.8\columnwidth]{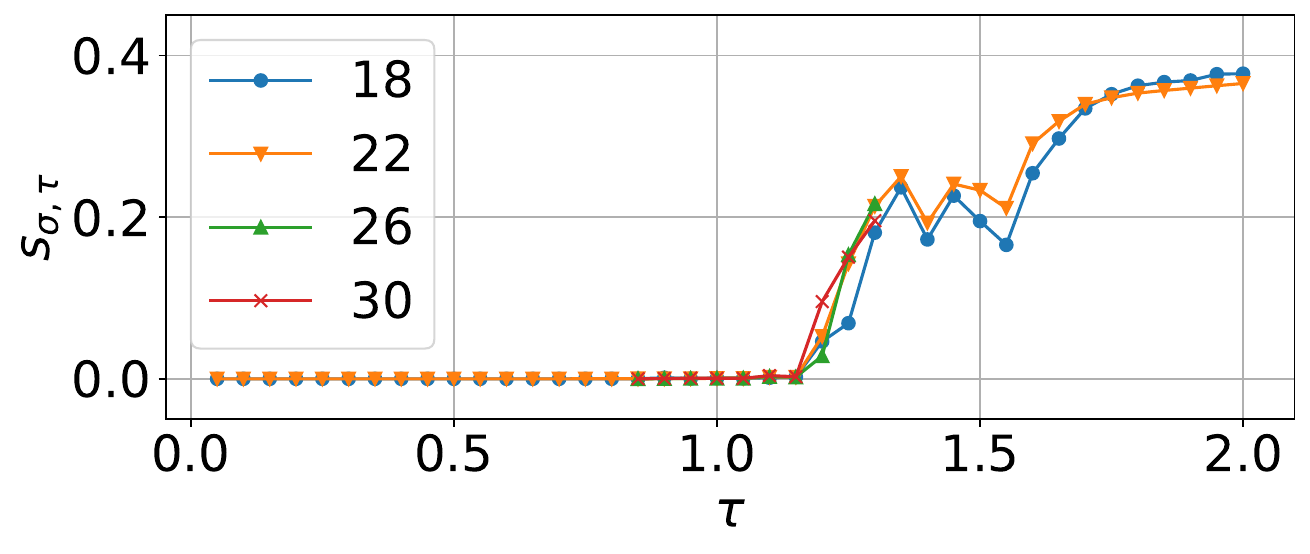}
	\caption{
        System-size dependence of time-averaged fidelity~\eqref{eq:time_averaged_fidelity_sigma}. Here the initial state is the ground state of $H_F^{(2)}$ and the cutoff is $\sigma=5\times10^3$.
    }
	\label{fig:Ldep}
\end{figure}

It is noteworthy that the threshold value $\tau\approx1.2$ does not correspond to the known convergence radius of the Magnus expansion.
Although some radii are known, each dictates that the expansion is convergent if $(\|H_1\|+\|H_2\|)\tau < r$, where $\|\cdots\|$ denotes the 2-norm, and $r$ is a constant, e.g., $r=\pi$~\cite{Blanes2009}.
Considering that both $\|H_1\|$ and $\|H_2\|$ are $O(L)$, we learn that those convergence radii are $O(L^{-1})$ and shrink as $L$ increases.
On the contrary, the transition-like behavior of the fidelity is robust for larger systems.
Figure~\ref{fig:Ldep} compares the fidelities at different system sizes at a fixed time cutoff $\sigma=5\times10^3$.
In the small-$\tau$ regime $\tau<1.2$, we observe little dependence on $L$, suggesting the stability of the Floquet ground state in the thermodynamic limit.
We note that $s_{\sigma,\tau}$ shows a small bump at $\tau=1.1$, visible on the log scale. This bump likely reflects an accidental many-body resonance (see Supplemental Material). We observe that it weakens with increasing system size such that our numerical results are consistent with the scenario that it vanishes in the thermodynamic limit $L\to \infty$.

\begin{figure}
    \includegraphics[width=0.85\columnwidth]{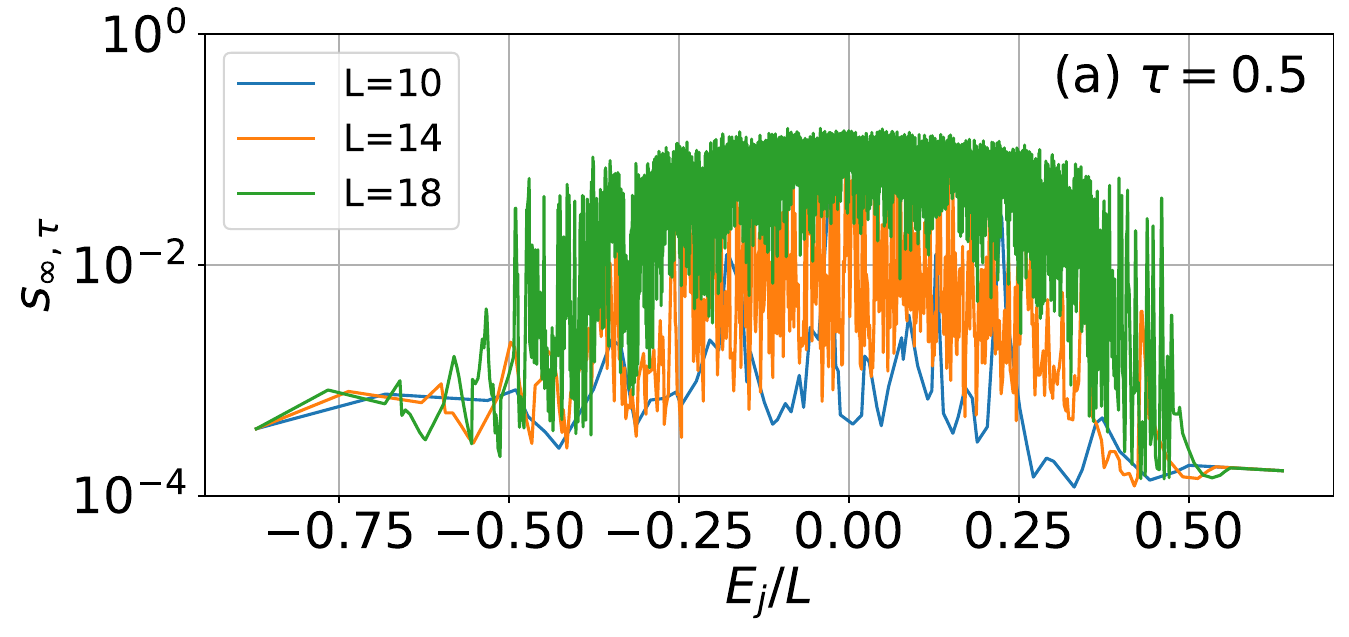}
     \includegraphics[width=0.85\columnwidth]{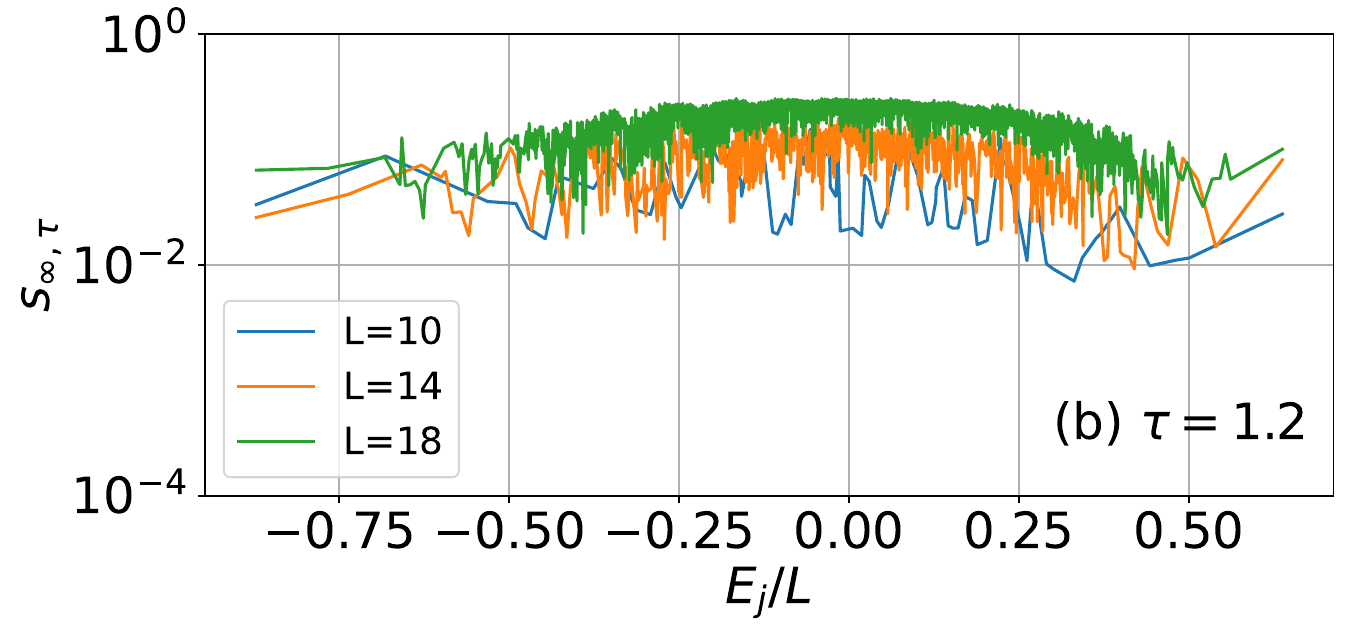}
	\caption{Scaled long-time fidelity error~\eqref{eq:def_s} for each eigenstate $\ket{\psi_\mathrm{ini}}=\ket{E_j}$ calculated by Eq.~\eqref{eq:time_averaged_fidelity}.
     The Trotter steps are (a) $\tau=0.5$ and (b) 1.2.
     Each color distinguishes the system sizes as in the legends.
    }
	\label{fig:F_eigen}
\end{figure}

Before looking into the interplay of the system size and the time cutoff, we examine other eigenstates of $H_F^{(2k)}$ chosen for the initial state.
For computational convenience, we calculate the infinite-time average $\overline{F}_{\infty,\tau}$ for each eigeneigenstate $\ket{E_j}$ of $H_F^{(0)}$, i.e., 
\begin{align}\label{eq:eigenkets}
H_F^{(0)}\ket{E_j}=E_j\ket{E_j}
\end{align}
for $j=0,\dots,d-1$.
Considering the translation and inversion symmetries shared by $T(\tau)$ and $H_F^{(0)}$, we restrict ourselves to the symmetry sector of the zero momentum and the even parity that hosts the ground state, and $d$ denotes the dimension of this symmetry sector.
Using the eigenstates of $T(\tau)$, 
\begin{align}
    T(\tau)\ket{\theta_\alpha} &= e^{-i\theta_\alpha} \ket{\theta_\alpha}
\end{align}
and assuming there is no degeneracy in the eigenvalues, we obtain
\begin{align}\label{eq:time_averaged_fidelity}
    \overline{F}_{\infty,\tau} = \sum_{\alpha}  |\braket{\theta_\alpha|E_j}|^4;
    \quad s_{\infty,\tau}=-L^{-1}\ln (\overline{F}_{\infty,\tau}).
\end{align}
Figure~\ref{fig:F_eigen} shows $s_{\infty,\tau}$ for all $\ket{\psi_\mathrm{ini}}=\ket{E_j}$ $(j=0,\dots,d-1)$.
The panel (a) is for $\tau=0.5$ below the crossover, where we observe that the GS ($j=0$), as well as the highest-excited state (HES, $j=d-1$), are consistent with the behavior $s_{\infty,\tau}={\rm const}(L)\ll 1$.
Note that this stability holds after the infinite Floquet cycles, without finite cutoff $\sigma$, at least up to $L\le 18$.
In contrast, $s_{\infty,\tau}$ tends to increase in the middle of the spectrum as $L$ increases, meaning the faster than exponential decay of fidelity with the system size.
For $\tau=1.2$ in the crossover, on the other hand, $s_{\infty,\tau}$ increases with $L$ for all states, including the GS and HES.
These results highlight the uniqueness of the GS and HES and potentially a few more nearby states as compared to other eigenstates. Namely the sharp crossover in fidelity (and other heating measures) seen for the GS in Fig.~\ref{fig:transition} is not present for generic initial eigenstates of $H_F^{(2k)}$. This finding seems rather unexpected as, naively, the notion of the ground state is not well defined for the Floquet unitary.

A physical interpretation for the unexpected robustness of the GS (and HES) is based on the breakdown of Fermi's golden rule (FGR) description of Floquet heating~\cite{Mallayya2019} as microscopic transitions among a continuum of eigenstates (see Supplemental Material for the supporting data and analysis).
First, we note that our unitary evolution~\eqref{eq:Trotter} is a special case of general Floquet evolution generated by a time-periodic Hamiltonian $H(t)=H_0+\epsilon g(t)K$.
In fact, Eq.~\eqref{eq:Trotter} realizes when $H_0=(H_1+H_2)/2$, $g(t)=\mathrm{sgn}(\cos(\Omega t))$ with $\Omega=\pi/\tau$, $K=(H_0-H_1)/2$, and $\epsilon=1$.
At small driving amplitudes $\epsilon$, the matrix elements $|\braket{E_{j'}|K|E_0}|^2$ excite the ground state into excited states at resonance $E_{j'}\approx E_0+\Omega$. Importantly, we are considering $L$-independent $\Omega$, and, in a standard scenario of local models, such lowest excitations $\ket{E_{j'}}$ can be viewed as a subextensive number of noninteracting quasiparticles with infinite lifetimes~\cite{SachdevBook}. In other words generic quantum systems at low energies behave as if they are free. As the system size increases this picture becomes more and more accurate. However, the noninteracting systems with bounded quasiparticle spectrum cannot absorb energy beyond their single or two-particle bandwidth. This, for example, follows from exponential, not factorial, operator spreading in the Krylov space~\cite{Avdoshkin2020}. This suppression of heating does not work for most excited states, i.e. if the initial energy $E_j$ is $O(L)$ above the ground energy, because it transitions into $E_{j'}$ with $E_{j'}\approx E_j+\Omega$ and these are dense and do not have quasiparticle description.
Interestingly, our analysis shows that this picture remains qualitatively unchanged if the driving amplitude $\epsilon$ is not small and the matrix elements of the unitary $\delta U=U_{2k}^\dag(\tau)T(\tau)$ replace the matrix elements of $K$ in the FGR analysis~\cite{Ikeda2021}. This robustness of the FGR threshold follows from the fact that within one driving period, the operators do not have time to spread, and hence $\delta U$ remains a quasi-local operator at any $\epsilon$.


\begin{figure}
    \includegraphics[width=0.8\columnwidth]{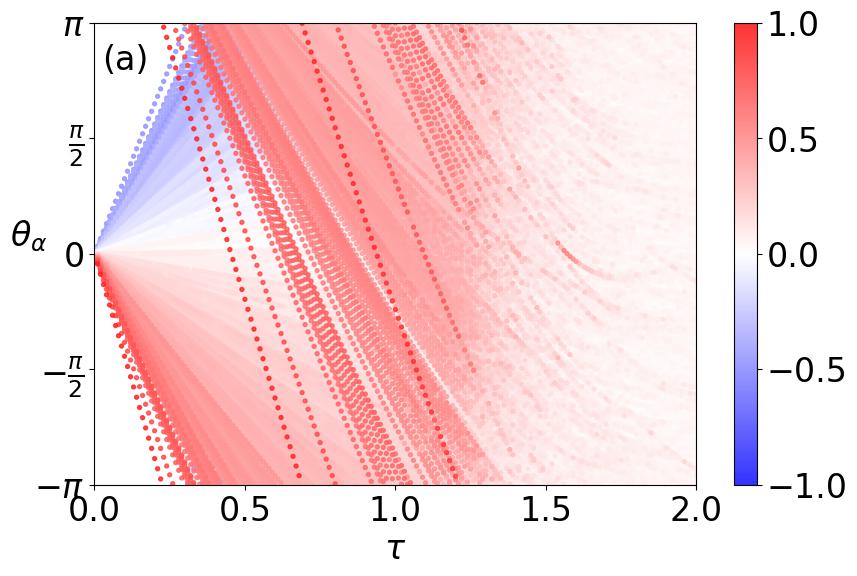} 
    \includegraphics[width=0.8\columnwidth]{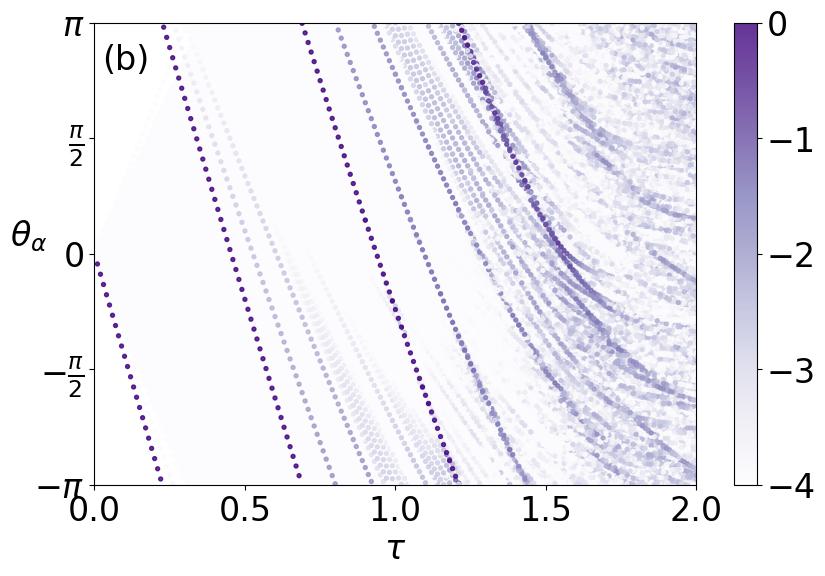} 
    \caption{All the $\theta_\alpha$'s plotted for various $\tau$ at system size $L=16$.
    Color of each data point shows (a) the magnetization expectation value $\braket{\theta_\alpha|m_z|\theta_\alpha}$
    and (b) the overlap with $H_F^{(0)}$'s GS, $\log_{10}|\braket{E_0|\theta_\alpha}|^2$.
    In panel (b), all the values less than $-4$ correspond to white color.}
    \label{fig:scar}
\end{figure}

\lsec{Approximate quantum many-body scars}
Now we return to the numerical analysis of the properties of the Floquet GS.
Equation~\eqref{eq:time_averaged_fidelity} dictates that the time-averaged fidelity is governed by the overlap between the initial state and the Floquet eigenstates.
Thus, the robustness of the Floquet ground state suggests the presence of a special eigenstate of the unitary $T(\tau)$, which is similar to the ground state of a local static Hamiltonian. We visualize the Floquet eigenstates $\ket{\theta_\alpha}$ in Fig.~\ref{fig:scar}(a), where we plot all the eigenvalues $\theta_\alpha$ for various $\tau$. These eigenvalues are color-coded according to their average magnetization
$\braket{\theta_\alpha|m_z|\theta_\alpha}$ with $m_z = L^{-1}\sum_{i=1}^L Z_i$.

A perturbation theory from $\tau=0$ allows us to interpret Fig.~\ref{fig:scar}(a) for small $\tau$.
Namely, we regard $U_0=e^{-iH_F^{(0)}\tau}$ as the unperturbed operator and $V := T(\tau)-e^{-iH_F^{(0)}\tau}=O(\tau^3)$ as the perturbation.
The eigenstates of $H_F^{(0)}$ defined in Eq.~\eqref{eq:eigenkets} satisfy $U_0\ket{E_j}=e^{-iE_j \tau}$, meaning $\ket{\theta_{j}}\approx \ket{E_{j}}$ with $\theta_j \equiv E_j\tau\ (\!\!\!\mod 2\pi)$ at the zeroth order.
If $\tau$ is so small that $|E_j|\tau<\pi$ holds for every $j$, the modulo can be ignored, and
the relations $\theta_j =E_j\tau$ $(j=0,1,\dots,d-1)$ are seen in Fig.~\ref{fig:scar}(a) for $\tau\lesssim0.2$.
Here the lowest (highest) branch of data is connected to the GS (HES) of $H_F^{(0)}$ in $\tau\to0^+$.

Once some of $|E_j|\tau$ exceeds $\pi$ as $\tau$ increases, the eigenvalues $\theta_\alpha$ are folded into the interval $[-\pi,\pi)$, which start to happen at $\tau\gtrsim 0.3$ (see Fig.~\ref{fig:scar}(a)).
The first folding occurs when $\|H_F^{(0)}\|\tau=\pi$, i.e., $\tau=O(L^{-1})$, which scales with $L$ like the convergence radius of the Magnus expansion.
After the folding, there appear pairs of eigenvalues of $U_0(\tau)$: $(e^{-i E_{j_1} \tau},e^{-i E_{j_2}\tau})$ coming closer, around which the perturbation $V$ is expected to hybridize the corresponding eigenstates.
Such hybridization should manifest as repulsion of magnitude $\sim |\braket{E_{j_1}|V|E_{j_2}}|$ between them.

Nevertheless, the ground-state and a few low-energy-state branches are robust even after the folding occurs, and the eigenvalues come across numerous other eigenvalues up to $\tau\approx1.2$.
Strictly speaking, there are level repulsions (see Supplementary Material), but these are very small and difficult to observe without fine-tuning $\tau$ (see also discussions below).
This is consistent with the expectation that the off-diagonal elements $\braket{E_{j_1}|V|E_{j_2}}$ are exponentially small in $|E_{j_1}-E_{j_2}|$, according to the off-diagonal eigenstate thermalization hypothesis (ETH)~\cite{Srednicki1999,DAlessio2016}.
We regard the robust eigenstates in the middle of the spectrum as the approximate many-body scar states since they are weakly mixed with the other states without any symmetry protection.
Similar robustness also exists near the HES branch, which becomes visible in reverse magnetization color-code plotting (see Supplemental Material).

The important role of the robust branches of Floquet eigenstates is shown in Fig.~\ref{fig:scar}(b),
which is a similar plot with the color being replaced by the overlap with the effective Floquet ground state, $|\braket{E_0|\theta_\alpha}|^2$ (in the log scale).
As the figure shows, only the ground-state branch is significantly populated for $\tau\lesssim0.5$,
a few other branches become gradually populated for $0.5\lesssim \tau \lesssim1.2$,
and the population scatters between states finally for $\tau\gtrsim1.2$.
These behaviors correspond to the smooth decrease of the fidelity (i.e., the increase of $s_{\sigma,\tau}$) in Fig.~\ref{fig:transition}(a) up to $\tau<1.2$
and its abrupt change in $\tau>1.2$.
A similar argument also holds when $\ket{E_0}$ is replaced by the GS of $H_F^{(2)}$, for which the overlap is concentrated even more on the GS branch.

\lsec{Role of the time cutoff $\sigma$}
Finally, we turn to $\sigma<\infty$ and discuss its role in removing spiky behaviors in $\overline{F}_{\infty,\tau}$ due to tiny level repulsions.
Note that
\begin{align}\label{eq:Fbar_finite}
    \overline{F}_{\sigma,\tau} = \sum_\alpha |c_\alpha|^4 + 2\sum_{\alpha < \beta} |c_\alpha|^2|c_\beta|^2 D_\sigma(\theta_\alpha-\theta_\beta),
\end{align}
where $c_\alpha = \braket{\theta_\alpha|\psi_\mathrm{ini}}$ and $D_\sigma(x)= (2\mathcal{N}_\sigma)^{-1} (1+\sum_{n=-\infty}^\infty e^{-(n/\sigma)^2 +in x})$, and $D_\sigma(x)$ has narrow peaks at $x=2\pi \mathbb{Z}$ of width $\sim\sigma^{-1}$ (see Supplemental Material for more details).
If we take the limit $\sigma\to\infty$ first with $L<\infty$ fixed, the second term in Eq.~\eqref{eq:Fbar_finite} vanishes and this equation reduces to Eq.~\eqref{eq:time_averaged_fidelity}.
As this expression changes significantly at each level repulsion, $\overline{F}_{\infty,\tau}$ becomes spiky.

The opposite order of limits, i.e., $L\to\infty$ first and $\sigma\to\infty$ second, would avoid the issue originating from the level repulsions, allowing one to study the thermodynamic limit for the shorter periods $\tau\lesssim1.2$.
This is because the second term in Eq.~\eqref{eq:Fbar_finite} can eliminate the spiky behavior of the first term if $\sigma^{-1}$ is larger than the tiny resonant level splittings, which exponentially go down with $L$ (see Supplemental Material for a more detailed discussion of these resonances). The numerical results are consistent with the scenario that if we take this order of limits then we can still observe a very sharp crossover as shown in Figs.~\ref{fig:transition} and \ref{fig:Ldep}, while at the same time eliminating the contribution of the accidental resonant spikes.

\lsec{Conclusions and Outlook}
We have shown strong evidence for a very sharp crossover or possibly even a phase transition in the long-time stability under a Floquet drive or Trotterized dynamics.
This transition has been well characterized using the effective Floquet ground state and taking the appropriate order of limits: $L\to\infty$ and then $\sigma\to\infty$.
Despite the common belief that all states eventually heat up to the infinite temperature under generic nonintegrable Floquet models, our results suggest that there are exceptional states with anomalously low or possibly zero Floquet heating above a critical driving frequency even in the thermodynamic limit.
Such states can be very interesting from the point of view of Floquet engineering because they are extremely long-lived.
We interpret the exceptional robustness of the Floquet ground state as a result of emergent free quasi-particle description of the low-energy states of local Hamiltonians with non-extensive energy $\Omega$. Such free quasi-particle systems with a bounded spectrum cannot absorb energy at driving frequencies above few-particle bandwidth.

There remain several open questions.
In particular, what is the class of Floquet models where such states exist and how does the number of such stable states scale with the system size~\cite{Evrard2023}? Can we find these states in the classical Floquet systems in the thermodynamic limit? Do such stable states exist for other, non Floquet, driving protocols?
In the extended-space picture~\cite{Sambe1973}, where the Floquet drive is represented by coupling to a static photon mode, the Floquet ground states can be viewed as stable mid-energy states, as they appear in the middle of the spectrum of the extended static Hamiltonian. As such, they appear to be similar to stable mid-energy states discovered in certain static models~\cite{Kormos2017,Castro-Alvaredo2020,Castro-Alvaredo2021}. It remains open to establish a precise correspondence with these models.
Of course it would be very interesting to see such stable states in experimental systems, such as nitrogen-vacancy centers in diamonds and ultracold atoms/ions. This transition as a function of the Trotter step size could also be observed in digital quantum simulators.

\lsec{Acknowlegements}
Fruitful discussions with Souvik Bandyopadhyay, Marin Bukov, Pieter Claeys, Isaac L. Chuang, Ceren B. Dag, Iliya Esin, Michael Flynn, Asmi Haldar, Martin Holthaus, Pavel Krapivsky, Takashi Oka, Tibor Rakovszky, and Dries Sels are gratefully acknowledged.
Numerical exact diagonalization in this work has been performed with the help of the QuSpin package~\cite{Weinberg2017,Weinberg2019}, and quantum time evolution with Qulacs~\cite{Suzuki2021}.
T. N. I. was supported by JST PRESTO Grant No. JPMJPR2112 and by JSPS KAKENHI Grant No. JP21K13852. A.P. was supported by NSF Grant DMR- 2103658 and the AFOSR Grant FA9550-21-1-0342.


\newpage
\onecolumngrid
\vspace{0.5cm}
\begin{center}
{\large \bf Supplemental Material: Robust effective ground state in a nonintegrable Floquet quantum circuit}
\end{center}

\setcounter{figure}{0}
\setcounter{equation}{0}
\setcounter{section}{0}
\renewcommand{\theequation}{S\arabic{equation}}
\renewcommand{\thesection}{S\arabic{section}}
\renewcommand{\thefigure}{S\arabic{figure}}

\section{S1. High-frequency expansion for fidelity}

The time-averaged fidelity (see Fig.~1 of the main text) shows smooth dependence on $\tau$ in the localized phase. Let us now show that quantitatively $1-\overline{F}_{\sigma,\tau}\propto \tau^{4k+4}$ with $k=0$ and 1 in the upper and lower panels, respectively.
This dependence directly follows from the combination of the high-frequency expansion and standard perturbation theory.

To derive this scaling we regard
$H_F^{(2k)}$ as the unperturbed Hamiltonian with the following eigenbasis:
\begin{align}\label{eq:eigenketsHF}
H_F^{(2k)}\ket{E_j^{(2k)}}=E_j^{(2k)}\ket{E_j^{(2k)}}.
\end{align}
In the non-heating phase we anticipate that the high-frequency expansion converges or almost converges such that there exists an accurate local approximation to the Floquet Hamiltonian $H_F$, which can be obtained using the BCH formula:
\begin{align}
\label{eq:BCH_Floquet}
    H_F = H_F^{(2k)} + V;\qquad V=O(\tau^{2k+2}).
\end{align}
This Hamiltonian must share eigenstates $\ket{\theta_\alpha}$ with the unitary $T(\tau)$. Note that at finite $L$ Eq.~\eqref{eq:BCH_Floquet} is guaranteed to converge for short enough $\tau$. Since $T(\tau)$ and $H_F$ share their eigenvectors, we aim to obtain $H_F$'s eigenvectors and eigenvalues,
\begin{align}
    H_F\ket{\mathcal{E}_j} =  \mathcal{E}_j \ket{\mathcal{E}_j},
\end{align}
which give $\bra{\theta_\alpha}\mathcal{E}_j\rangle\approx \delta_{\alpha j}$ and $\theta_\alpha = \mathcal{E}_\alpha\tau\;{\rm mod}\; 2\pi$.
Then the leading-order perturbation theory~\cite{Messiah2014} gives the overlaps $\braket{E_\alpha|\mathcal{E}_\beta}$.

Now we consider the time-averaged fidelity [Eq.~(10) in the paper] with the initial state being $\ket{E_0^{(2k)}}$ and assume that $\sigma$ is so large that $D_\sigma(E_0\tau - E_j\tau)$ are negligibly small for all $j\neq0$.
Then we have
\begin{align}
 s_{\sigma,\tau}= -L^{-1}\ln \left( \overline{F}_{\sigma,\tau}\right) \approx -L^{-1}\ln\left(|\braket{E_0^{(2k)}|\mathcal{E}_0}|^4 \right)
    \approx 2L^{-1} \sum_{j\neq0} 
    \frac{ |\braket{ E_0^{(2k)} |V| E_j^{(2k)} }|^2 }{(E_0^{(2k)} - E_j^{(2k)})^2}=O(\tau^{4k+4}).
\end{align}
While the formal derivation of this result requires convergence of the BCH (Magnus) expansion, we observe that this scaling works very well in the entire localized phase $\tau\lesssim 1.2$.

\section{S2. Many-body resonances}
\begin{figure}
         \includegraphics[width=0.6\columnwidth]{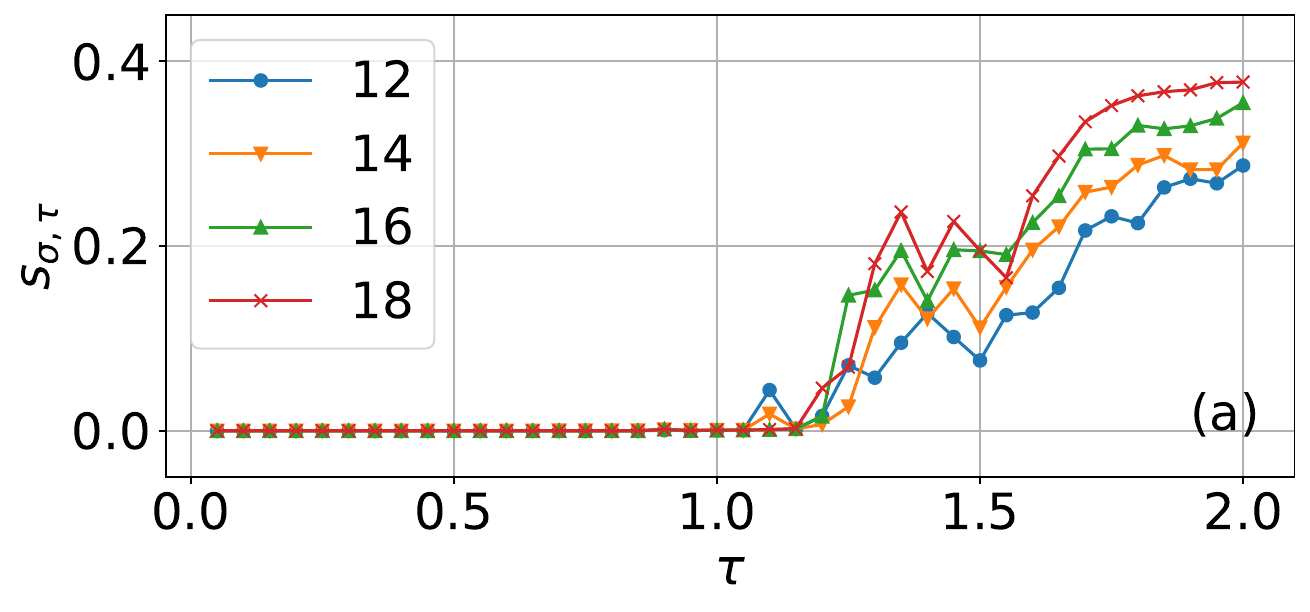}\\
         \includegraphics[width=0.48\columnwidth]{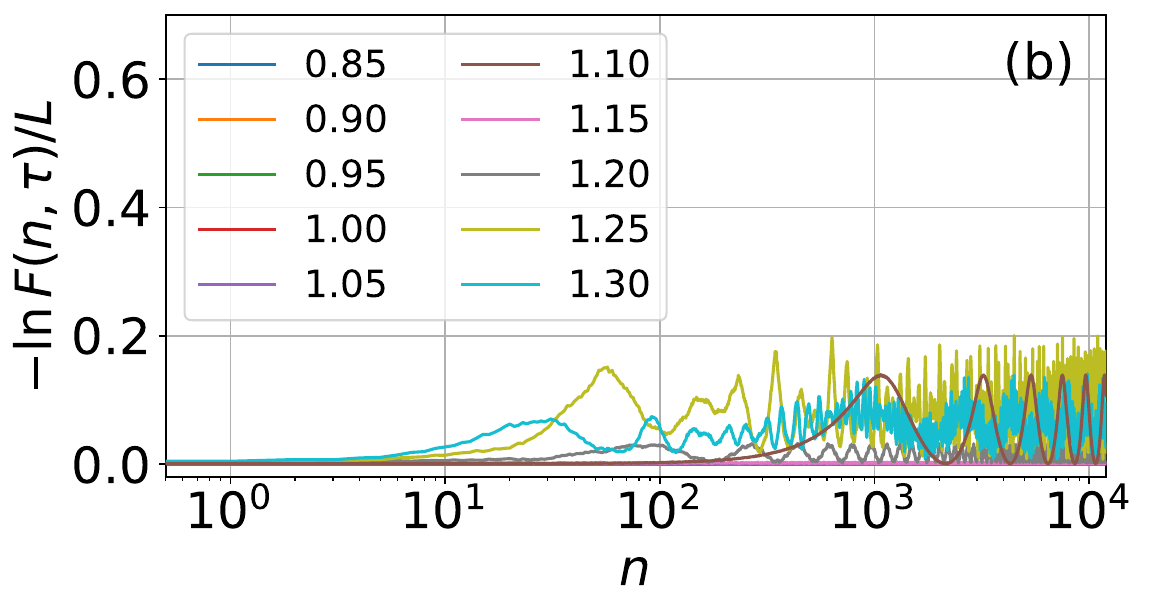}
         \includegraphics[width=0.48\columnwidth]{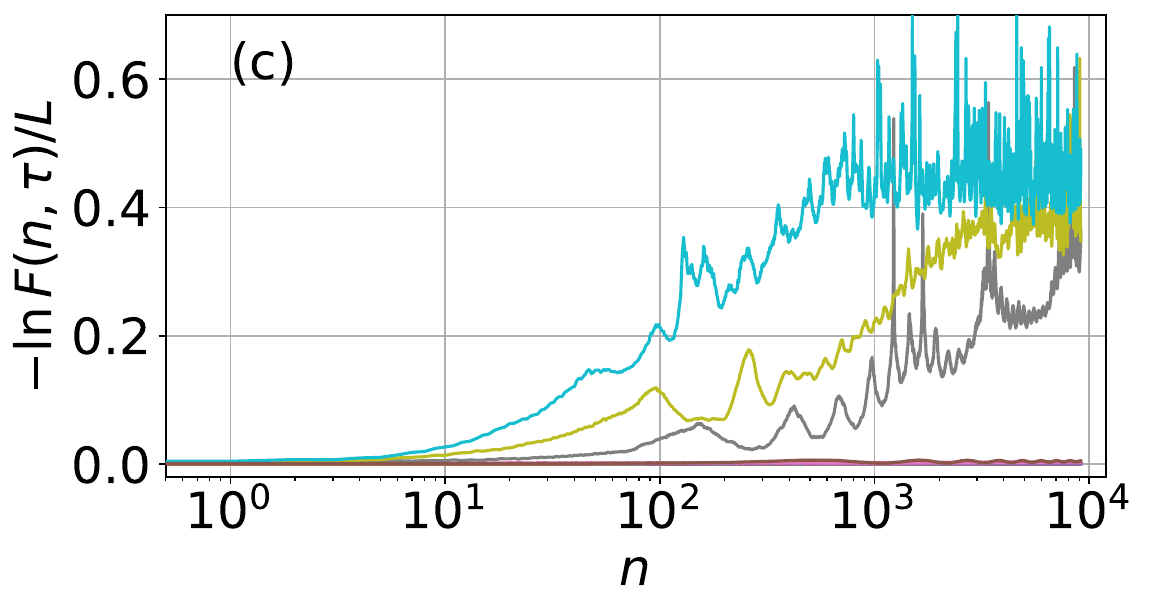}
	\caption{
 (a) System-size dependence of Time-averaged fidelity. Here the initial state is the ground state of $H_F^{(2)}$, the cutoff is $\sigma=5\times10^3$, and the system sizes are shown in the legends.
 (b,c) Dynamics of fidelity at (b) $L=12$ and (c) 30 with the initial state being the ground state of $H_F^{(2)}$.
        The legends on the central panel show the values of $\tau$ and apply to the right panel.}
	\label{figS:resonance}
\end{figure}

Here we show supplemental data in the vicinity of $\tau=1.2$, around which the time-averaged fidelity exhibits an abrupt change.
Figuire~\ref{figS:resonance}(a) extends Fig.~2 in the main text to smaller systems down to $L=12$.
At $L=12$, we observe a bump, which diminishes as $L$ increases.
To take a closer look at this bump, we show the explicit time dependence of fidelity at $L=12$ in Fig.~\ref{figS:resonance}(b).
Here we observe a large persistent sinusoidal oscillation at $\tau=1.1$.
This type of oscillation implies that, when we expand our initial state $\ket{\psi_\mathrm{ini}}$ in terms of the Floquet eigenstate $\ket{\theta_\alpha}$ as
$\ket{\psi_\mathrm{ini}} = \sum_\alpha c_\alpha\ket{\theta_\alpha}$,
two of them are dominantly populated: $\ket{\psi_\mathrm{ini}}\approx c_{\alpha_1}\ket{\theta_{\alpha_1}}+c_{\alpha_2}\ket{\theta_{\alpha_2}}$.
In fact, if this is the case, we have $F(n,\tau) = |\braket{\psi_\mathrm{ini}| T(\tau)^n |\psi_\mathrm{ini}}|^2 \approx \left| |c_{\alpha_1}|^2 e^{-i\theta_{\alpha_1}n} + |c_{\alpha_2}|^2 e^{-i\theta_{\alpha_2}n}\right|^2 
\approx \left| 1 -|c_{\alpha_1}|^2 \sin^2 (\frac{\theta_{\alpha_1}-\theta_{\alpha_2}}{2}n) \right|^2 $, where we used $|c_{\alpha_1}|^2+|c_{\alpha_2}|^2\approx 1$.
We recall that, for most small $\tau$, $\ket{\psi_\mathrm{ini}}$ is dominated by a single Floquet eigenstate, and the fidelity is approximately constant.
Then it is natural to expect that the Floquet eigenstate can become hybridized with another accidentally (i.e. through an isolated many-body resonance like in Ref.~\cite{Bukov2016heating}). This happens around $\tau=1.1$, where our initial state has a significant weight on both Floquet eigenstates.

In the heating regime one can anticipate that with increasing $L$ these resonances proliferate leading to eventual monotonic in $\tau$ decay of fidelity. This is indeed the case for excited states of $H_F^{(2)}$. For the ground state, conversely, our numerical results are more consistent with the scenario, where this resonance fades away for larger systems (see Fig.~\ref{figS:resonance}(c)). For the largest accessible system size $L=30$, the amplitude of the fidelity oscillation at $\tau=1.1$ is strongly suppressed, as Fig.~\ref{figS:resonance}(c) illustrates.
It is worth noting that this panel shows that the lifetime of the initial state discontinuously increases as $\tau$ decreases below $\tau=1.2$.
This discontinuous change underlies the abrupt jump of the time-averaged fidelity shown in Fig.~1 in the paper.

\section{S3. Additional details of the Floquet spectrum}
\begin{figure}
    \includegraphics[width=0.5\columnwidth]{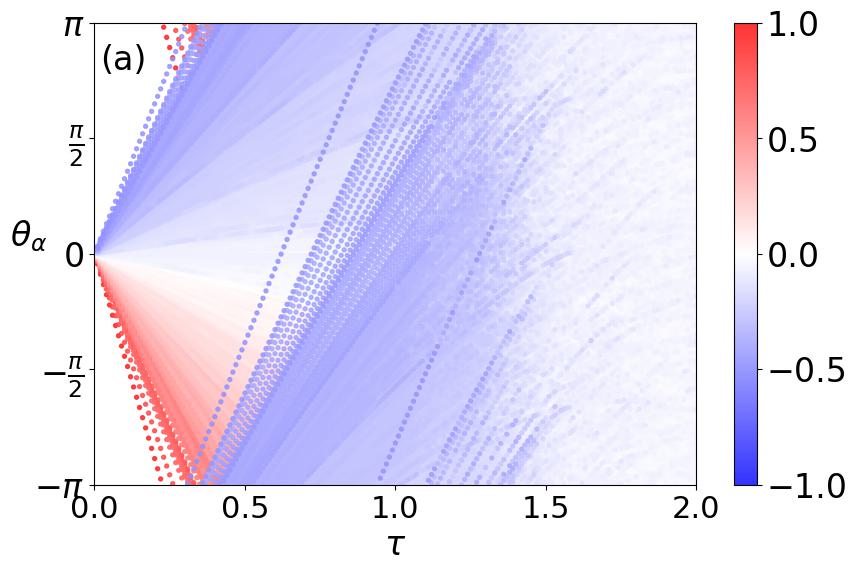} \\
    \includegraphics[width=0.45\columnwidth]{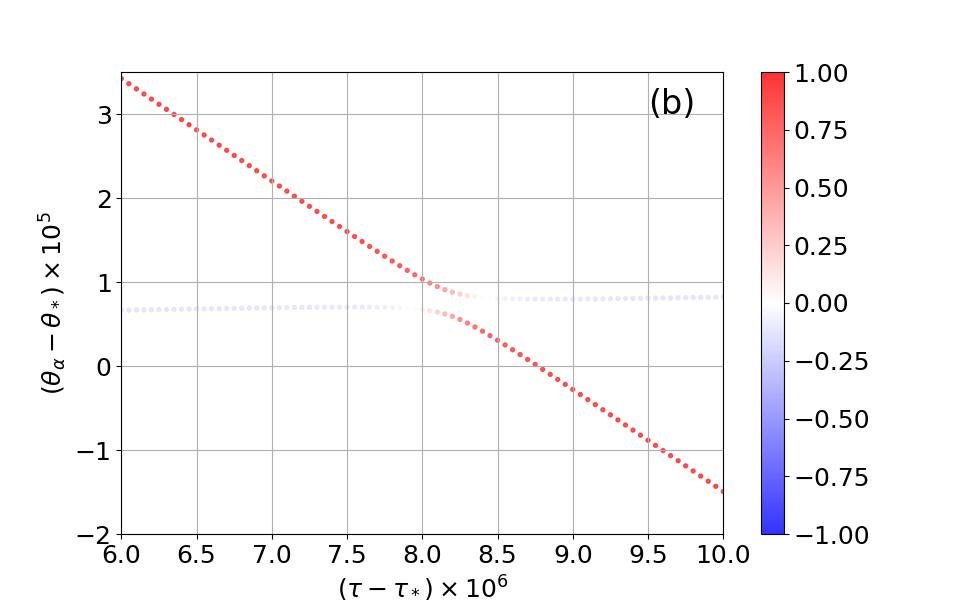} 
    \includegraphics[width=0.45\columnwidth]{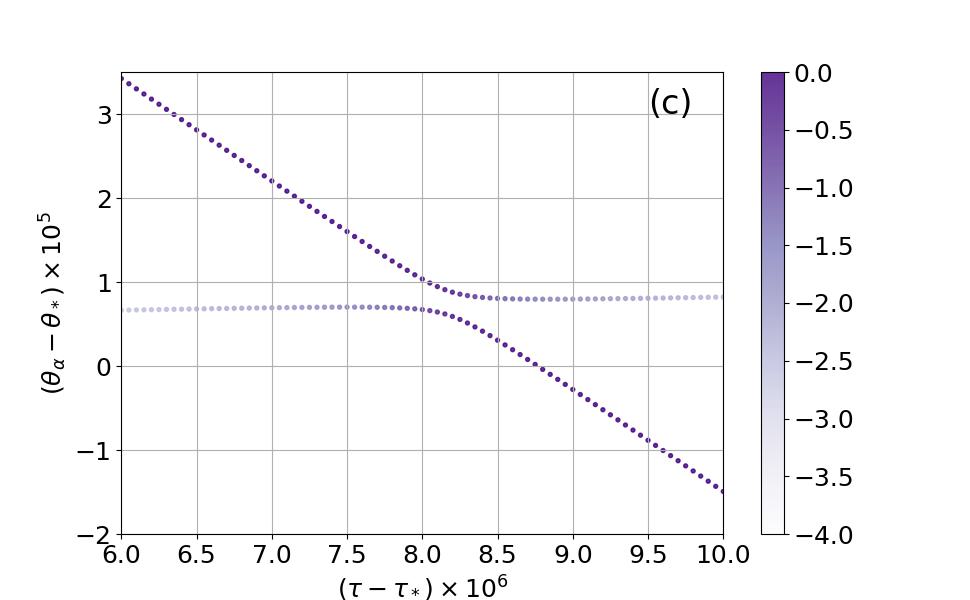} 
    \caption{
    (a) All the $\theta_\alpha$'s plotted for various $\tau$ at system size $L=16$.
    Color of each data point shows the magnetization expectation value $\braket{\theta_\alpha|m_z|\theta_\alpha}$.
    In contrast to Fig.~4, eigenstates with smaller expectation values are plotted more in front.
    (b,c) Magnified eigenvalue plots around $\tau_*=0.89995$ and $\theta_*=0.4570$.
    Colors show (b) the magnetization expectation value and (c) the overlap with $H_F^{(0)}$'s GS, $\ln_{10}|\braket{E_0|\theta_\alpha}|^2$, as the color bars indicate.
    }
    \label{figS:scar}
\end{figure}

In Fig.~4(a), the color coding is chosen such that the eigenstates with larger expectation values of the magnetization are emphasized. This scheme allows us to emphasize the Floquet ground state. In Figure~\ref{figS:scar}(a) we emphasize states with negative magnetization, bringing them to the front. In this way we can visualize the states close to the HES, allowing us to observe the robustness of the HES branch.

Figures~\ref{figS:scar}(b,c) show the structure of the level crossing of the special Floquet GS with one of the generic eigenstates. The plot region is fine-tuned to the vicinity of one of such crossings. The robust GS branch shows tiny hybridization in the narrow width $\delta \tau\approx 10^{-6}$, as expected from the absence of symmetry protecting this eigenstate exactly. Away from such tiny regions, the GS branch is robust dominating the overlap with the states $\ket{E_0^{(0)}}$ and $\ket{E_0^{(2)}}$.
If we literally follow one of eigenstate adiabatically from $\tau=\tau_*-\delta\tau$ to $\tau_*+\delta\tau$, it will become a distinct state (red to blue in Fig.~\ref{figS:scar}(b)).
This is a version of the absence of the adiabatic limit~\cite{Hone1997,Holthaus2015} and has been noticed and well-studied in single-particle models.

Such a tiny repulsion makes $\overline{F}_{\infty,\tau}$ spiky as a function of $\tau$.
As in Eq.~(9) in the main text, the two eigenstats, which we call $\ket{\theta_{\alpha_0}}$ and $\ket{\theta_{\beta_0}}$, contribute to the fidelity as $|\braket{\theta_{\alpha_0}|E_0}|^4+|\braket{\theta_{\beta_0}|E_0}|^4$.
For $\tau\gtrsim \tau_\ast +\delta \tau$ and $\tau\lesssim \tau_\ast-\delta \tau$,
only one of these terms dominates such that the total contribution of these two states to fidelity is approximately the same.
However, in the vicinity of the repulsion, i.e., $|\tau-\tau_*|\lesssim \delta \tau$, $|\braket{\theta_{\alpha_0}|E_0}|^4$ and  $|\braket{\theta_{\beta_0}|E_0}|^4$ come close and their sum decreases significantly.
This spiky behavior smears out if we choose $\sigma$ to be smaller than the inverse resonance width $1/\min(\theta_{\alpha_0}-\theta_{\beta_0})$.

\section{S4. Model deformation and connection with the Fermi's golden rule}
In order to study the connection of the Floquet ground state with the standard linear response theory, we introduce a new parameter $\epsilon$, which plays the role of the driving amplitude.
The deformed Hamiltonian is
\begin{align}
    H_\epsilon(t) = H_0 + \epsilon g(t)K;\quad
    H_0 = \frac{H_1+H_2}{2};\quad
    K = \frac{H_1-H_2}{2},
\end{align}
\begin{align}\label{eq:gt}
    g(t) = \begin{cases}
        1 & \left( k\le t/\tau < k+\frac{1}{2} \right) \\
        -1 & \left( k+\frac{1}{2} \le t/\tau < k+\frac{3}{2} \right)\\
        1 & \left(  k+\frac{3}{2} \le t/\tau < k+2 \right),
    \end{cases}
\end{align}
with $k\in\mathbb{Z}$ and $2\tau$ is the driving period. Note that the time averaged Hamiltonian $H_0=H_F^{(0)}/2$ does not depend on $\epsilon$. Clearly this model reduces to the original one at $\epsilon=1$. When $\epsilon\ll 1$, the external driving $\epsilon g(t)K$ can be considered as a perturbation, and Fermi's golden rule (FGR) is expected to approximate the heating well~\cite{Mallayya2019,Mallayya2019x}.
For $\epsilon\neq \pm1$, the one-cycle unitary
\begin{align}\label{eq:deformedU}
U=e^{-i (H_0+\epsilon K)\frac{\tau}{2}} e^{-i (H_0-\epsilon K)\tau}e^{-i (H_0+\epsilon K)\frac{\tau}{2}}
\end{align}
is not in a Trotterized form, and the efficient circuit simulator cannot be used without further approximations, so we resort to smaller system sizes, where exact diagonalization is possible.

\begin{figure}
         \includegraphics[width=\columnwidth]{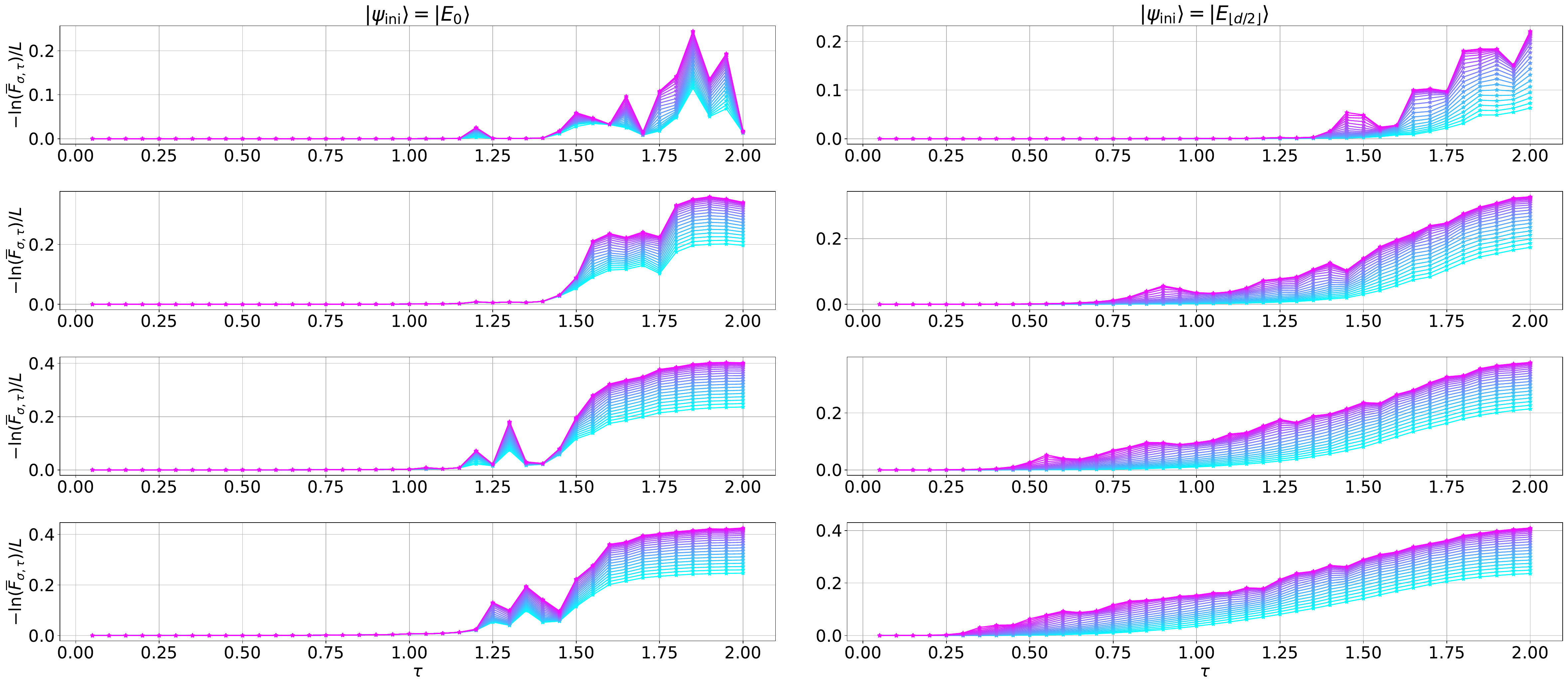}
	\caption{
        Time-averaged fidelity in the log scale under the Floquet evolution~\eqref{eq:deformedU} starting from the ground states of $H_0$ (left) and the mid-excited state, $j=\lfloor d/2\rfloor$ (right).
        The perturbation amplitude is $\epsilon=0.1$, 0.3, 0.5, and 0.75, from top to bottom.
        The system size is $L=18$, and plot colors from light blue to magenta correspond to the time cutoff $\sigma$ from $10^2$ to $10^4$ equidistant in the log scale.
    }
	\label{figS:fidelity}
\end{figure}

Figure~\ref{figS:fidelity} shows the counterparts of Fig.~1 in the paper obtained for the deformed model.
When the initial state is the ground state of $H_0$ (left panels), the time-averaged fidelity shows sharp crossovers as $\tau$ increases, except for accidental bumps, and the crossover is the sharpest for the largest perturbation $\epsilon=0.75$ in the panels. We thus see that the transition becomes most pronounced at large driving amplitudes consistent with observations of Ref.~\cite{Haldar2018}. At the same time we observe a strong qualitative difference between sharpness of transition between heating and no heating regimes between the initial ground state and generic excited states of $H_0$ for any value of $\epsilon$.

The difference between these two different initial states is also seen from the FGR viewpoint.
Here we implement the Floquet FGR~\cite{Ikeda2021}, which reduces to the conventional FGR~\cite{Mallayya2019,Mallayya2019x} for $\epsilon\ll1$, to compare the FGR prediction to the exact simulation.
Namely, neglecting the superposition of energy eigenstates, we consider the following master equation
\begin{align}\label{eq:master}
  \frac{d  P_j(t) }{dt} = \sum_{j'} [w_{j'\to j} P_{j'} (t)  - w_{j\to j'} P_j(t)],
\end{align}
where $P_j(t)$ represents the probability on $\ket{E_j}$ satisfying $\sum_j P_j(t) =1$, and the transition rate $w_{j\to j'}$ from $\ket{E_j}$ to $\ket{E_{j'}}$ is given by
\begin{align}\label{eq:FGRrates}
    w_{j\to j'} &
    = \frac{2\pi}{T^2}|\braket{E_{j'}| U| E_{j} }|^2 \sum_{l\in\mathbb{Z}} \delta( E_{j'} - E_{j} -l \Omega ) \\
    &\approx \frac{2\pi}{T^2}|\braket{E_{j'}| U| E_{j} }|^2 \sum_{l\in\mathbb{Z}}
    \frac{1}{\Delta\sqrt{\pi}} \exp\left( -\frac{( E_{j'} - E_{j} -l \Omega )^2}{\Delta^2} \right),
\end{align}
where $T=2\tau=2\pi/\Omega$, $E_j$ being eigenvalues of $H_0$, and $\Delta>0$ is a Gaussian regularization of the delta function.
If the initial state is an eigenstate of $H_0$ and $\braket{E_j | \psi_\mathrm{ini}}\propto \delta_{jj_0}$ holds for a $j_0$, $F(n,\tau)=[P_{j_0}(n\tau)]^2$, which is obtained by solving the master equation~\eqref{eq:master} from the initial condition $P_j(0)=\delta_{jj_0}$.
With this initial condition, we solve the master equation~\eqref{eq:master} and obtain the fidelity numerically.

\begin{figure}
         \includegraphics[width=\columnwidth]{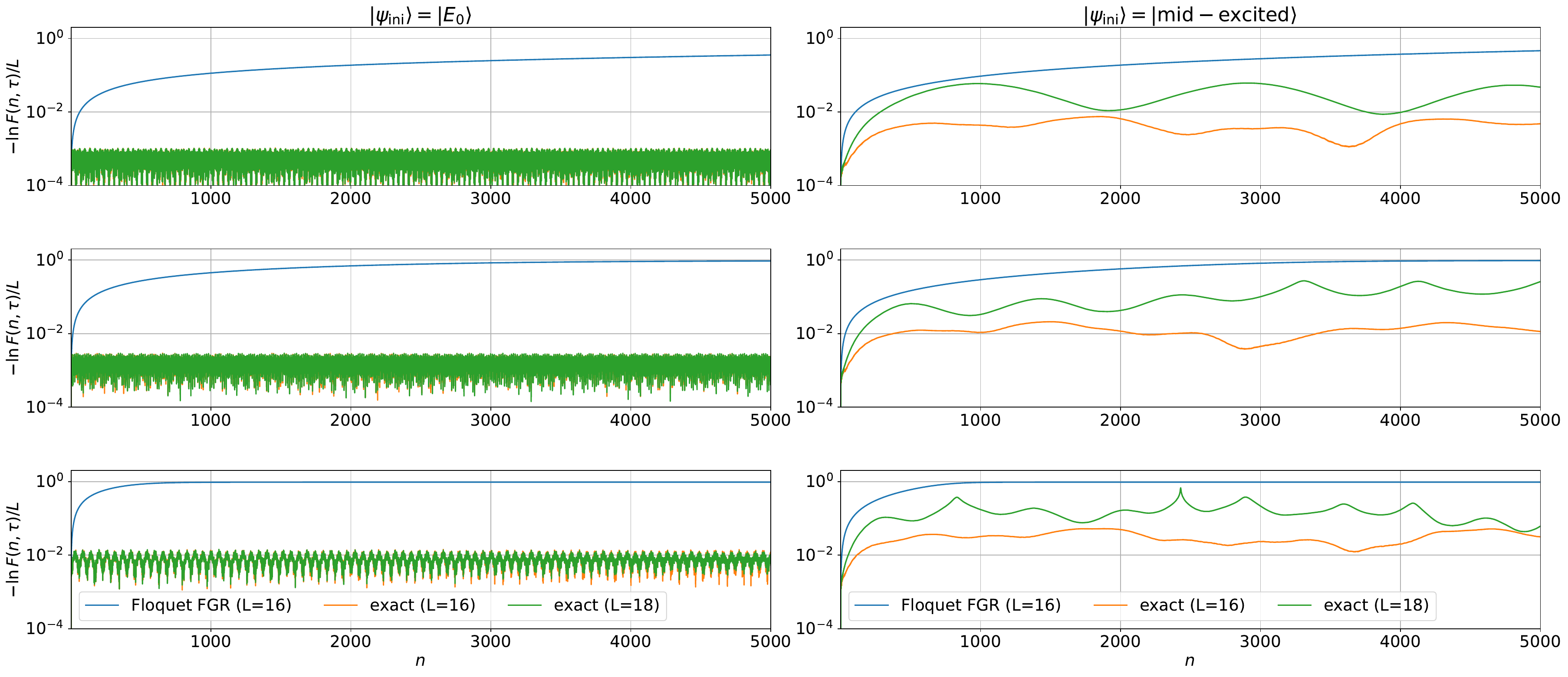}
	\caption{
        Time evolution of fidelity $F(n,\tau)$ calculated by the FGR for $L=16$ (blue) and the exact diagonalization of $U$ at $L=16$ (orange) and $L=18$ (green).
        Left and right panels correspond, respectively, to the cases where the initial state is the ground state and the mid-excited state whose eigenenergy is closest to zero.
        The driving amplitude is $\epsilon=0.3$ in all panels, and $\tau=0.9$, 1.1, and 1.3 from top to bottom.
    }
	\label{figS:FGR}
\end{figure}

\begin{figure}
         \includegraphics[width=0.6\columnwidth]{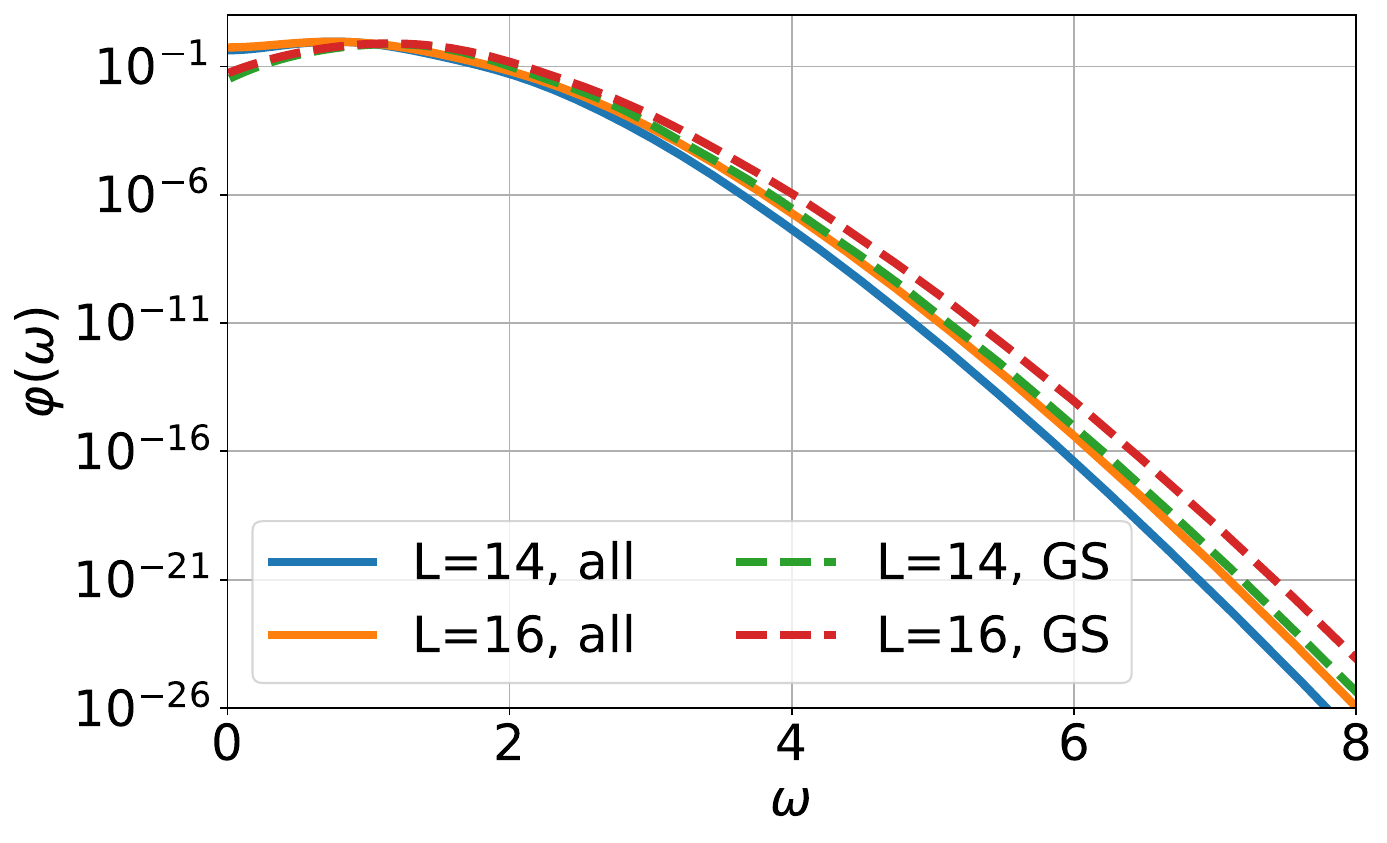}
	\caption{
    Spectral functions for the ground state $\varphi_0(\omega)$ (solid) and for the state average $\varphi(\omega)$ (dashed).
    }
	\label{figS:spec}
\end{figure}

Figure~\ref{figS:FGR} compares the FGR with the exact fidelity dynamics.
In the FGR calculation, we set $\Delta = 0.035 L$ like in Ref.~\cite{Ikeda2021}.
When the initial state is the mid-excited state whose eigenenergy is the closest to zero (right column), the FGR qualitatively captures the fidelity dynamics except for some finite-size effects. Note, however, the performance of the FGR for fidelity is quantitatively worse than for macroscopic observables~\cite{Ikeda2021} because the former is more sensitive to small errors in the wave function. Nevertheless it is clear that as the system size increases the FGR becomes more accurate. Conversely, when the initial state is the ground state, the FGR fails to capture the fidelity dynamics even qualitatively and the accuracy of the FGR does not improve with increasing $L$. These results suggest that the stability of the effective ground state is not  captured by the FGR. To double-check that this is indeed the case we numerically compute the spectral function of the operator $K$, which determines the FGR rate~\cite{Mallayya2019,Sels2021}:
\begin{align}
    \varphi_j(\omega) \equiv  \sum_{j'} |\langle E_{j'}| K | E_j \rangle |^2 \delta (E_{j'}-E_j -\omega)
    \approx 
     \sum_{j'} |\langle E_{j'}| K | E_j \rangle |^2
    \frac{1}{\Delta\sqrt{\pi}} \exp\left( -\frac{( E_{j'} - E_{j} - \omega )^2}{\Delta^2} \right).
\end{align}
We observe that it is qualitatively similar between $j=0$ and its average over all states:
\begin{align}
    \varphi(\omega) = D^{-1}\sum_j \varphi_j(\omega).
\end{align}
These are shown in Fig.~\ref{figS:spec}. Apart from a small dip in the ground state spectral function at low frequencies, which is due to the energy gap, we do not see any qualitative differences between the ground state and excited state spectral functions. We can thus conclude that the exceptional stability of the effective ground state must be related to the sparse resonant structure of the ground state spectral function not captured by the FGR. Such structure was discussed, in particular, in the context of disordered systems~\cite{Crowley_2022,Morningstar_2022}, although it was argued that this sparseness is a property of all states.

\section{S5. Time-dependent perturbation theory and emergence of the FGR threshold}

Here, we analyze in detail the mechanism behind the robustness of the Floquet ground state within the FGR heating mechanism. Because as we explained in the previous section the driving amplitude does not play a particularly important role, we return here to the original case of $\epsilon=1$. If the driving period $\tau$ is sufficiently small the effective Hamiltonian $H_F^{(2k)}$ well approximates the true unitary so that $\delta U \equiv U_{2k}(\tau)^\dag T(\tau)=e^{iH_F^{(2k)}\tau}T(\tau)\approx 1$.
Now we suppose our initial state is the $j$-th eigenstate of $H_F^{(2k)}$, i.e., $H_F^{(2k)}\ket{E_j^{(2k)}}=E_m^{(2k)}\ket{E_j^{(2k)}}$ and consider the fidelity in $n$ Floquet cycles
\begin{align}
    F_j(n,\tau) \equiv |\braket{E_j| T^n(\tau)|E_{j}}|^2.
\end{align}
Here and below we omit the order specification $(2k)$ for simplicity.
For $j=0$ (i.e., the ground state), $F_j$ reduces to the fidelity discussed in the main paper.
The fidelity $F_j$ can also be interpreted as the survival probability of the initial state $\ket{E_j}$.
We are comparing $F_j$ for $j=0$ and $j$ in the middle of the spectrum.
Using the normalization condition $1=\sum_{j'} |\braket{E_{j'}| T^n(\tau)|E_{j}}|^2.$ and introducing the transition probability to another $j'$-th eigenstate in $n$ cycles,
\begin{align}
    p_{j\to j'} \equiv |\braket{E_{j'}| T^n(\tau)|E_{j}}|^2 \qquad (j\neq j'),
\end{align}
we have 
\begin{align}
    -\ln [F_j(n,\tau)] = -\ln\left(1- \sum_{j'(\neq j)}p_{j\to j'}\right) \approx \sum_{j'(\neq j)}p_{j\to j'}.
\end{align}

As shown in Ref.~\cite{Ikeda2021}, we approximate $p_{j\to j'}$ in the leading order of $\delta U$, having
\begin{align}
    p_{j\to j'} \approx \left| \sum_{l=1}^n e^{i(\theta_j-\theta_{j'})l} \right|^2 |\braket{E_j|\delta U| E_{j'}}|^2
    =\frac{\sin^2 \left(\frac{\theta_j -\theta_{j'}}{2}n\right)}{\sin^2 \left(\frac{\theta_j -\theta_{j'}}{2}\right)}|\braket{E_j|\delta U| E_{j'}}|^2,
\end{align}
where $\theta_j \equiv E_j \tau$ and we assumed $e^{i(\theta_j-\theta_{j'})}\neq 1$.
Using this approximation, we obtain the following approximation for the fidelity 
\begin{align}\label{eq:Fapprox}
    -\ln [F_j(n,\tau)] \approx -\ln[\tilde{F}_j(n,\tau)] \equiv \sum_{j'(\neq j)}\frac{\sin^2 \left(\frac{\theta_j -\theta_{j'}}{2}n\right)}{\sin^2 \left(\frac{\theta_j -\theta_{j'}}{2}\right)}|\braket{E_j|\delta U| E_{j'}}|^2.
\end{align}

If the matrix elements $|\braket{E_j|\delta U| E_{j'}}|$ are smooth and nonzero functions of $E_i-E_j$ this expression leads to the standard FGR expression where $-\ln[\tilde{F}_j(n,\tau)]$ linearly increases with $n$. However, if the state $j$ is the ground state isolated from the continuum and there are no matrix elements for states $j'$ with $E_{j'}-E_j=\Omega$ like it happens in, for example, free particle systems, there are no resonant contributions with $\theta_{j'}-\theta_j=2\pi n\;\leftrightarrow\; E_{j'}-E_j=2\pi n \Omega$. In this case we can replace the fast oscillating term $\sin^2 \left(\frac{\theta_j -\theta_{j'}}{2}n\right)$ with its average $1/2$. Hence in the long time limit $n\to \infty$, the fidelity should be well approximated by its long time average:
\begin{align}\label{eq:FapproxInf}
-\ln[\tilde{F}_j(n,\tau)]\approx -\ln[\tilde{F}_j^\infty(n,\tau)] = \frac{1}{2}\sum_{j'(\neq j)} \frac{|\braket{E_j|\delta U| E_{j'}}|^2}{\sin^2 \left(\frac{\theta_j -\theta_{j'}}{2}\right)}\qquad (n\gtrsim \frac{1}{\Delta\theta_j}),
\end{align}
where $\Delta \theta_j$ stands for the typical level spacing for the state $\ket{E_j}$. Note that for extensive systems both $-\ln[F_j(n,\tau)]$ and $-\ln[\tilde{F}_j(n,\tau)]$ are expected to scale linearly with the system size so it is more convenient to analyze the fidelity densities or equivalently the rate functions such as $-L^{-1}\ln[F_j(n,\tau)]$.
 
\begin{figure}
         \includegraphics[width=\columnwidth]{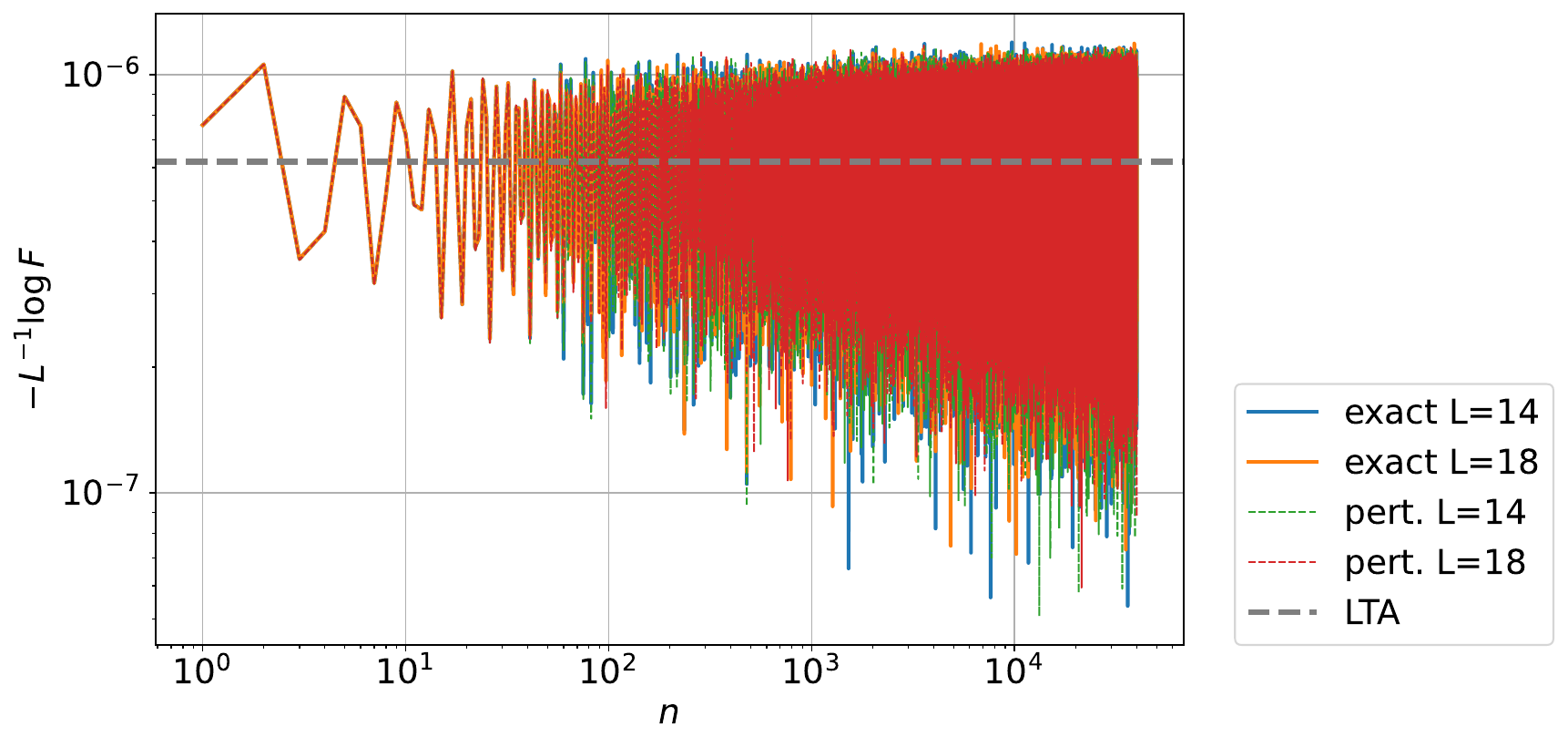}
         \includegraphics[width=\columnwidth]{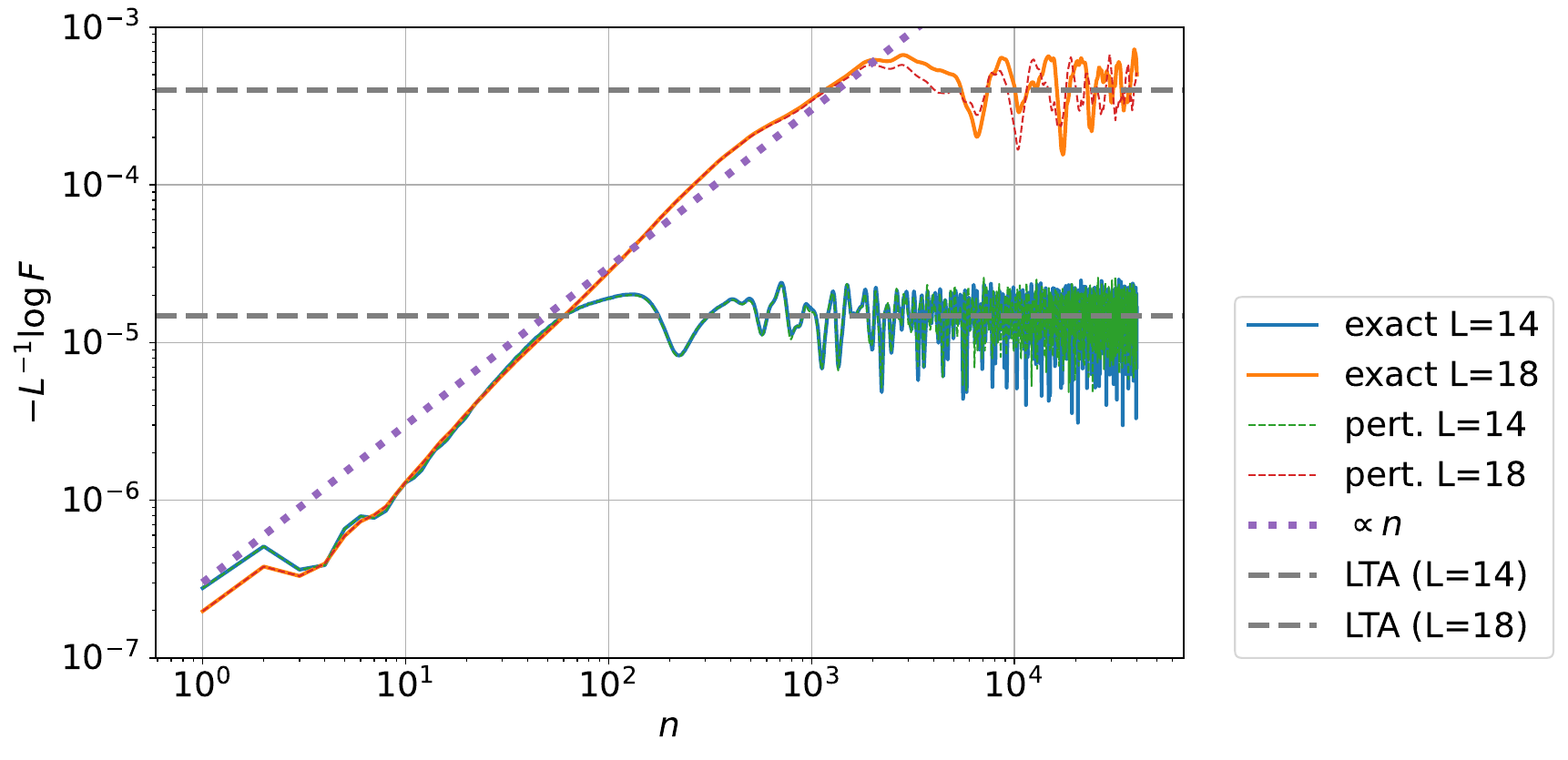}
 \caption{
     Scaled exact $F_j(n,\tau)$ (solid) and approximate $\tilde{F}_j(n,\tau)$ (dashed) fidelities for the ground state $j=0$ (upper panel) and a mid-excited state ($j$ with the minimum $|E_j|$) (lower panel).
    The driving period is $\tau=0.5$, and dotted lines in the lower panel are guides to the eye showing linear scaling in $n$.
     Thick dashed lines show the long-time average~\eqref{eq:FapproxInf}, which in the upper panel are on top of each other for $L=14$ and $18$.} 
	\label{figS:pert}
\end{figure}

Figure~\ref{figS:pert} compares the exact $F_j(n,\tau)$ and the approximate $\tilde{F}_j(n,\tau)$ fidelities for several system sizes at $\tau=0.5$ below the threshold.
Here and below we consider the second-order effective Hamiltonian $H_F^{(2k)}=H_F^{(2)}$.
In the figure, we show the results for the ground state $j=0$ and another state in the middle of the spectrum such that $|E_j|$ is the minimum for each $L$.
In both cases, the perturbation theory approximation $\tilde{F}_j(n,\tau)$ approximates the exact one almost perfectly except for the oscillation details after the saturation.
A difference between the two initial states is that the saturation value of the normalized fidelity $-L^{-1}\ln F_j(n,\tau)$ stays constant (increases) with $L$, indicating infinite temperature FGR-type heating for the excited states.

Notably, while the fidelity behaves very differently in time $n$ for the ground and excited states, the perturbation theory works quite well in both cases.
This implies for the excited states on can use the standard assumption in deriving the FGR replacing the sum over $j'$ in Eq.~\eqref{eq:Fapprox} by a continuous integral over $\theta$, with the help of $\int_{-\pi/2}^{\pi/2}d\theta \sin^2(n\theta)/\sin^2(\theta)\approx \pi n$ $(n\in\mathbb{Z})$. With the additional assumption that the relevant matrix elements $|\braket{E_{j'}|\delta U| E_{j}}|$ are positive and smooth functions of $\theta_{j'}$ we then arrive into the Floquet FGR expression for the linear log-fidelity growth with $n$. In contrast, for the ground state the resonances, where the matrix elements $|\braket{E_{j'}|\delta U| E_{j}}|$ are non-zero appear to be very sparse that we can instead set $\sin^2(n\theta)\approx 1/2$ and in this way obtain $n$-independent expression for the average fidelity plus additional $n$-dependent fluctuations.

\begin{figure}
         \includegraphics[width=\columnwidth]{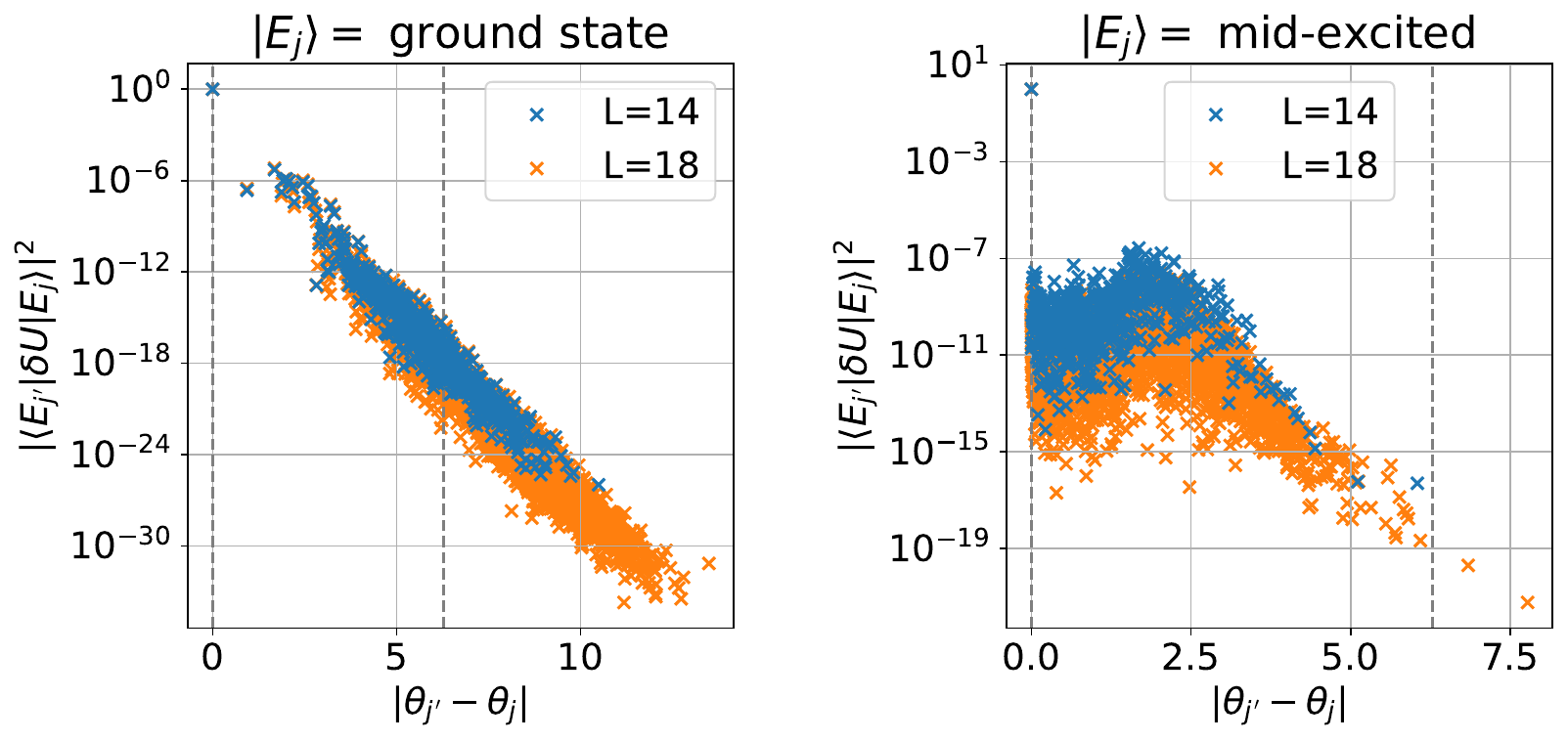}
 \caption{Squared matrix elements $|\braket{E_{j'}|\delta U|E_{j}}|^2$ for the reference state $\ket{E_j}$ corresponding to the ground state (left) and the mid-excited state (right). 
The vertical lines show $2\pi m$ ($m\in\mathbb{Z}$, and $\tau=0.5$)
    }
	\label{figS:Umat}
\end{figure}

To further analyze the perturbative fidelity in
Figure~\ref{figS:Umat} we show the matrix elements $|\braket{E_{j'}|\delta U|E_{j}}|^2$ for $j$ fixed to be the ground state and the mid-excited state, respectively.
For the mid-excited state, there are dense contributions from nearby degenerate states $\theta_j\approx \theta_{j'}$.
This also happens for most excited states that have dense nearby states on the spectrum.
For the ground state, instead, the resonant contribution comes from $\theta_{j'}\approx \theta_{0} + 2\pi$ because there is no $\theta_{j'}\approx \theta_{0}$ due to the energy gap above the ground state.
The resonant contribution is coming from $E_{j'} \approx E_{0} + \Omega$. We are interested in the limit where $\Omega$ is $L$-independent constant.

\begin{figure}
         \includegraphics[width=0.7\columnwidth]{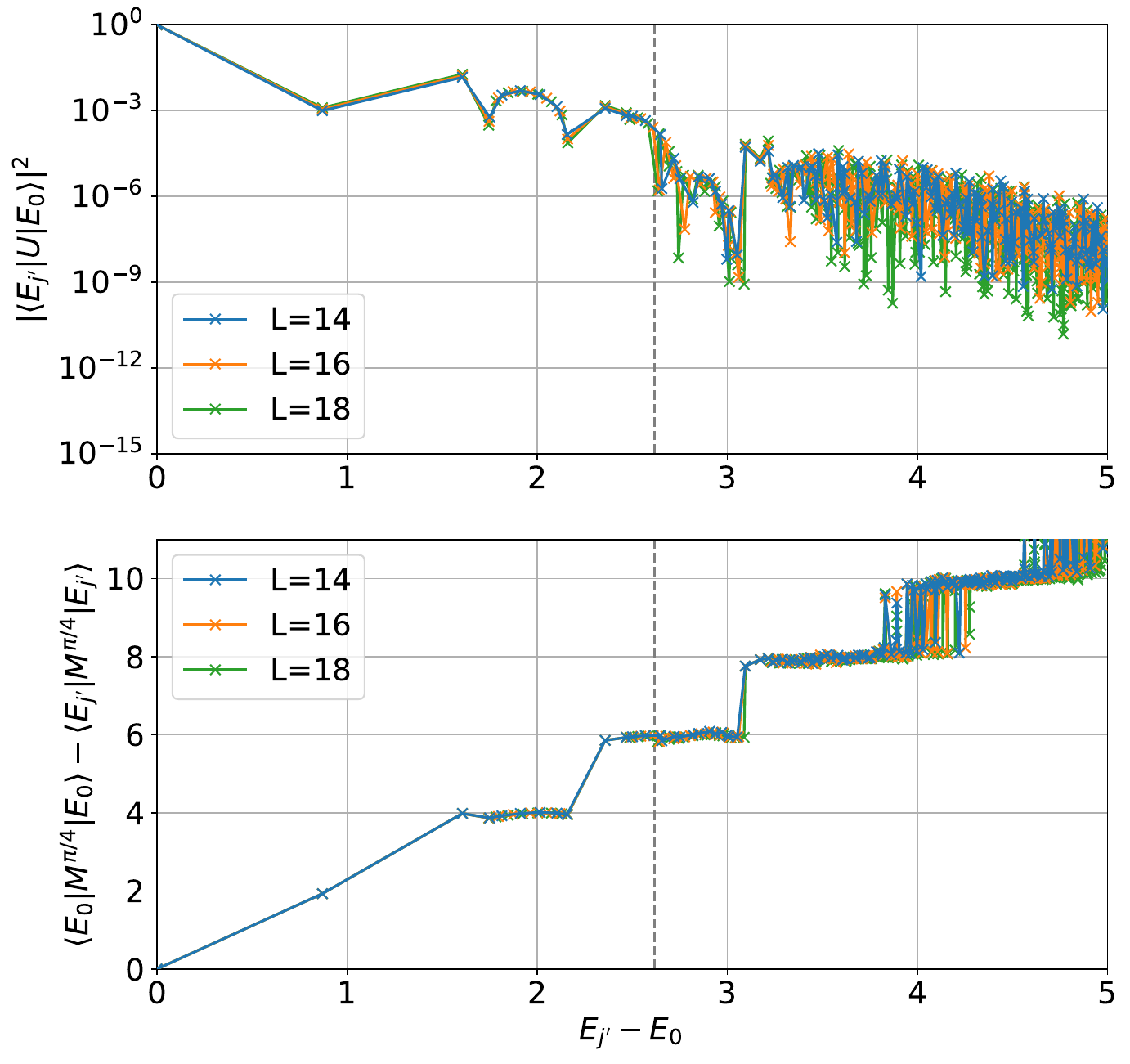}
 \caption{(top) Squared matrix elements $|\braket{E_{j'}|\delta U|E_{0}}|^2$ between each eigenstate of $H_F$ and the ground state.
 (bottom) The difference in the expectation values of the tilted magnetization $M^{\pi/4}$ between the ground state and each eigenstate of $H_F$.
 In both panels, the vertical lines show the driving frequency $\Omega=\pi/\tau$, and the period is fixed at $\tau=1.2$.
    }
	\label{figS:mag}
\end{figure}

\begin{figure}
         \includegraphics[width=\columnwidth]{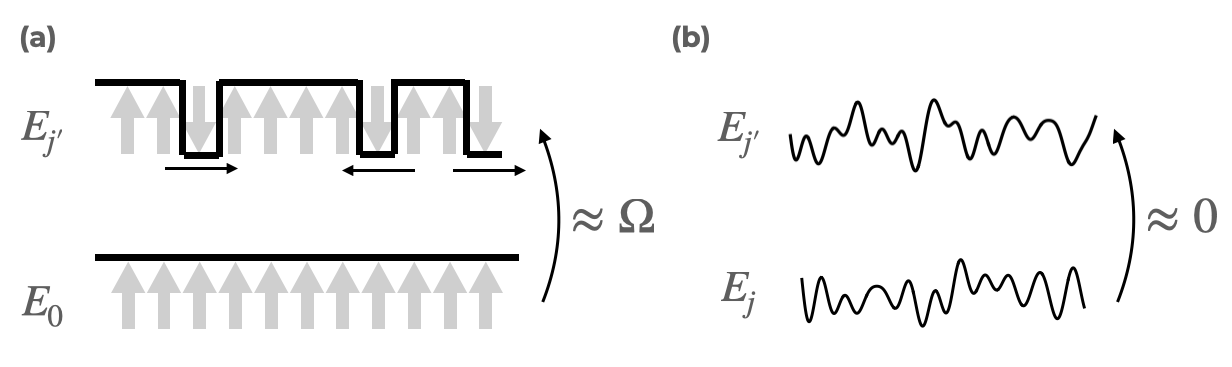}
 \caption{Schematic picture of dominant transitions from (a) the ground state and (b) high excited states.
 (a) The ground state transitions to states whose energy is $\Omega=O(1)$ energy above, and these states involve sub-extensive numbers of quasiparticles.
 (b) High excited states do not have quasiparticle descriptions, and transitions to dense degenerate states are possible.
     }
	\label{figS:excitation}
\end{figure}

In a standard scenario of local models, excitations above the ground state with low non-extensive energies can be viewed as isolated dressed quasi-particles with asymptotically infinite lifetimes in thermodynamic limit~\cite{SachdevBook}.
 In our effective Hamiltonian $H_F$, such quasi-particles are approximately local spin flips along the external field $-\frac{1}{2}(g,0,h)$ in the time-averaged Hamiltonian $H_0$ (recall that we set $g=h=1$).
To count these quasi-particles, we compute the magnetization along this field $M^{\pi/4}=\frac{1}{\sqrt{2}}\sum_{i=1}^L(\sigma^x_i+\sigma^z_i)$. The corresponding expectation values are plotted in the lower panel of Fig.~\ref{figS:mag}. The ground state is strongly polarized along the field and has a large expectation value of the magnetization, while the low excited states involve a few spin flips, each of which changes the magnetization by approximately 2.
While our numerical results are limited to finite systems where these quasi-particles necessarily interact, it is expected that as the system size increases (at a fixed $\Omega$) the independent quasi-particle structure becomes more and more accurate. So the excited states $|E_{j'}\rangle$ entering the perturbative fidelity~\eqref{eq:Fapprox} can be asymptotically described by those of a free noninteracting model. The physical picture highlighting the special role of the ground state and the low energy excitations is schematically illustrated in Fig.~\ref{figS:excitation}. For such free models, it is well understood that there is no energy absorption, at least within FGR above the single or a few-particle bandwidth (see e.g. Ref.~\cite{Pereira_2009}). This conclusion is also supported by the slow (exponential vs.~fast factorial) operator spreading in free models consistent with a finite absorption edge~\cite{Parker_2019}. Of course, in finite systems, these quasi-particles will interact with each other leading to accidental many-body resonances as indeed obvious from e.g. Fig.~\ref{figS:fidelity}. However, as the system size increases, the noninteracting picture should work better and better, and these resonances should gradually disappear in agreement with our numerical results. Note that this is exactly opposite to proliferation of the resonance with the system size for the excited states (see e.g. Ref.~\cite{Bukov2016heating}). Finally we note that this argument for a sharp heating threshold is perturbative. It indicates that the FGR heating rate sharply decreases beyond some self-consistently determined critical driving period. In Fig.~\ref{figS:mag}, the critical driving frequency $\Omega$ for $\tau=1.2$ is shown to be in the three-particle states, and the matrix elements $\braket{E_{j'}|\delta U|E_j}$ quickly drops at the frequency. As we change the driving period $\tau$ the spectrum of the effective Floquet Hamiltonian is only weakly affected. So the main effect of changing $\tau$ is that the dashed line indicating the resonant absorption energy moves through the spectrum. We cannot exclude existence of some other, for example, non-FGR heating mechanisms for such stable Floquet ground states. But even if such mechanism exists it should lead to anomalously slow heating. Interestingly, our numerical analysis of the FGR rate does not rely on the smallness of the driving amplitude $\epsilon$ and is rather relies on smallness and locality of the unitary $\delta U$ as it follows from validity of the perturbative expression~\eqref{eq:Fapprox}.

\section{S6. Properties of $D_\sigma(x)$}
\begin{figure}
         \includegraphics[width=0.4\columnwidth]{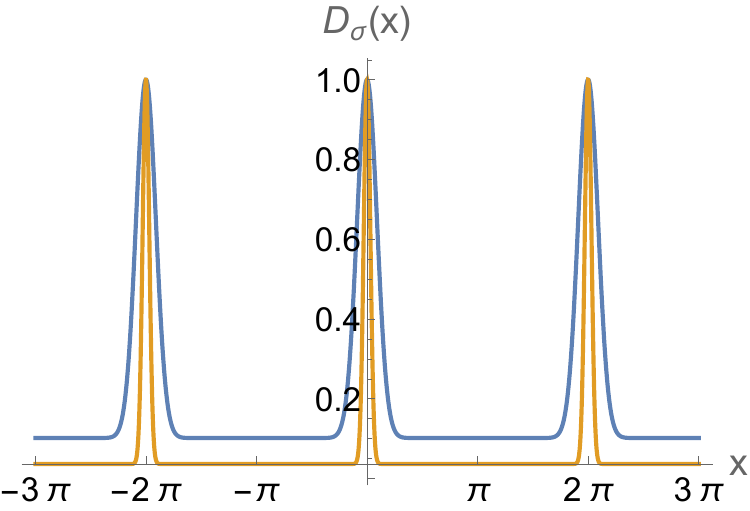}
	\caption{
    The function $D_\sigma(x)$ for $\sigma=5$ (blue) and 15 (orange).
    }
	\label{figS:theta}
\end{figure}

Here we supplement the properties of
\begin{align}
D_\sigma(x)= (2\mathcal{N}_\sigma)^{-1} \left(1+\sum_{n=-\infty}^\infty e^{-(n/\sigma)^2 +in x}\right)
=\frac{1+\sum_{n=-\infty}^\infty e^{-(n/\sigma)^2 +in x}}{1+\sum_{n=-\infty}^\infty e^{-(n/\sigma)^2}}.
\end{align}
Invoking the Jacobi theta function~\cite{theta}
\begin{align}
    \vartheta_3(z|\tau) \equiv \sum_{n=-\infty}^{\infty} e^{i \pi \tau n^2}e^{2 n i z}
\end{align}
and substituting $\tau=i (\pi \sigma^2)^{-1}$ and $z=0,x/2$, we obtain
\begin{align}
    D_\sigma(x) = \frac{1+\vartheta_3(\frac{x}{2}| \frac{i}{\pi \sigma^2})}{1+\vartheta_3(0| \frac{i}{\pi \sigma^2})}.
\end{align}
Properties of the theta function~\cite{theta} ensures the periodicity $D_\sigma(x+2\pi)=D_\sigma(x)$ and $D_\sigma(2\pi l)=1$ for $l\in \mathbb{Z}$.
Also, $D_\sigma(x)$ rapidly decays as $x$ deviates from $2\pi l$ like $\sim \exp(-\sigma^2 (x-2\pi l)^2/4)$, and its minimum scales as $\propto \sigma^{-1}$.
Figure~\ref{figS:theta} shows typical behaviors of $D_\sigma(x)$.

\twocolumngrid

\end{document}